\documentclass[10pt,fleqn]{article}
\usepackage{graphics,epsfig}
\begin{document}

\renewenvironment{description} {\list{}{ \itemsep 1mm \parsep 0mm
    \itemindent -10mm \leftmargin 10mm }} {\endlist}
\newcommand{\bgm}[1]{\mbox{\boldmath $#1$}}
\newcommand{\bgms}[1]{\mbox{{\scriptsize \boldmath $#1$}}}
\newcommand{\ul}[1]{\underline{#1}} \newcommand{\bgt}[1]{{\boldmath
    $#1$}}


\centerline{\Large\bf Predictability, complexity and learning}
\bigskip\bigskip

\centerline{\large William Bialek,$^1$ Ilya Nemenman,$^{1,2}$ and
  Naftali Tishby$^{1,3}$} \bigskip

\centerline{$^1$NEC Research Institute, 4 Independence Way,
  Princeton,
  New Jersey 08540} \centerline{$^2$Department of Physics, Princeton
  University, 
  Princeton, New Jersey 08544} \centerline{$^3$School of
  Computer Science and Engineering, 
  and Center for Neural Computation,} \centerline{ Hebrew
  University, Jerusalem 91904, Israel} \bigskip

\centerline{\today}

\bigskip\bigskip\hrule\bigskip\bigskip

We define {\em predictive information} $I_{\rm pred} (T)$ as the
mutual information between the past and the future of a time series.
Three qualitatively different behaviors are found in the limit of
large observation times $T$: $I_{\rm pred} (T)$ can remain finite,
grow logarithmically, or grow as a fractional power law. If the time
series allows us to learn a model with a finite number of parameters,
then $I_{\rm pred} (T)$ grows logarithmically with a coefficient that
counts the dimensionality of the model space. In contrast, power--law
growth is associated, for example, with the learning of infinite
parameter (or nonparametric) models such as continuous functions with
smoothness constraints.  There are connections between the predictive
information and measures of complexity that have been defined both in
learning theory and in the analysis of physical systems through
statistical mechanics and dynamical systems theory.  Further, in the
same way that entropy provides the unique measure of available
information consistent with some simple and plausible conditions, we
argue that the divergent part of $I_{\rm pred} (T)$ provides the
unique measure for the complexity of dynamics underlying a time
series.  Finally, we discuss how these ideas may be useful in
different problems in physics, statistics, and biology.

\vfill\newpage
\tableofcontents

\vfill\newpage

\section{Introduction}

There is an obvious interest in having practical algorithms for
predicting the future, and there is a correspondingly large literature
on the problem of time series extrapolation.\footnote{The classic
  papers are by Kolmogoroff (1939, 1941) and Wiener (1949), who
  essentially solved all the extrapolation problems that could be
  solved by linear methods. Our understanding of predictability was
  changed by developments in dynamical systems, which showed that
  apparently random (chaotic) time series could arise from simple
  deterministic rules, and this led to vigorous exploration of
  nonlinear extrapolation algorithms (Abarbanel et al. 1993).  For a
  review comparing different approaches, see the conference
  proceedings edited by Weigend and Gershenfeld (1994).}  But
prediction is both more and less than extrapolation: we might be able
to predict, for example, the chance of rain in the coming week even if
we cannot extrapolate the trajectory of temperature fluctuations. In
the spirit of its thermodynamic origins, information theory (Shannon
1948) characterizes the potentialities and limitations of all possible
prediction algorithms, as well as unifying the analysis of
extrapolation with the more general notion of predictability.
Specifically, we can define a quantity---the {\em predictive
  information}---that measures how much our observations of the past
can tell us about the future.  The predictive information
characterizes the world we are observing, and we shall see that this
characterization is close to our intuition about the complexity of the
underlying dynamics.

Prediction is one of the fundamental problems in neural computation.
Much of what we admire in expert human performance is predictive in
character---the point guard who passes the basketball to a place where
his teammate will arrive in a split second, the chess master who knows
how moves made now will influence the end game two hours hence, the
investor who buys a stock in anticipation that it will grow in the
year to come.  More generally, we gather sensory information not for
its own sake but in the hope that this information will guide our
actions (including our verbal actions). But acting takes time, and
sense data can guide us only to the extent that those data inform us
about the state of the world at the time of our actions, so the only
components of the incoming data that have a chance of being useful are
those that are predictive. Put bluntly, {\em nonpredictive information
  is useless to the organism}, and it therefore makes sense to isolate
the predictive information. It will turn out that most of the
information we collect over a long period of time is nonpredictive, so
that isolating the predictive information must go a long way toward
separating out those features of the sensory world that are relevant
for behavior.

One of the most important examples of prediction is the phenomenon of
generalization in learning.  Learning is formalized as finding a model
that explains or describes a set of observations, but again this is
useful only because we expect this model will continue to be valid: in
the language of learning theory [see, for example, Vapnik (1998)] an
animal can gain selective advantage not from its performance on the
training data but only from its performance at generalization.
Generalizing---and not ``overfitting'' the training data---is
precisely the problem of isolating those features of the data that
have predictive value (see also Bialek and Tishby, in preparation).
Further, we know that the success of generalization hinges on
controlling the complexity of the models that we are willing to
consider as possibilities. Finally, learning a model to describe a
data set can be seen as an encoding of those data, as emphasized by
Rissanen (1989), and the quality of this encoding can be measured
using the ideas of information theory.  Thus the {\em exploration of
  learning problems should provide us with explicit links among the
  concepts of entropy, predictability, and complexity}.

The notion of complexity arises not only in learning theory, but also
in several other contexts.  Some physical systems exhibit more complex
dynamics than others (turbulent vs.\ laminar flows in fluids), and
some systems evolve toward more complex states than others (spin
glasses vs.\ ferromagnets).  The problem of characterizing complexity
in physical systems has a substantial literature of its own; for an
overview see Bennett (1990).  In this context several authors have
considered complexity measures based on entropy or mutual information,
although as far as we know no clear connections have been drawn among
the measures of complexity that arise in learning theory and those
that arise in dynamical systems and statistical mechanics.

An essential difficulty in quantifying complexity is to distinguish
complexity from randomness.  A true random string cannot be compressed
and hence requires a long description; it thus is complex in the sense
defined by Kolmogorov (1965, Li and Vit{\'a}nyi 1993, Vit{\'a}nyi and
Li 2000), yet the physical process that generates this string may have
a very simple description.  Both in statistical mechanics and in
learning theory our intuitive notions of complexity correspond to the
statements about complexity of the underlying process, and not
directly to the description length or Kolmogorov complexity.

Our central result is that {\em the predictive information provides a
  general measure of complexity} which includes as special cases the
relevant concepts from learning theory and from dynamical systems.
While work on complexity in learning theory rests specifically on the
idea that one is trying to infer a model from data, the predictive
information is a property of the data (or, more precisely, of an
ensemble of data) itself without reference to a specific class of
underlying models.  If the data are generated by a process in a known
class but with unknown parameters, then we can calculate the
predictive information explicitly and show that this {\em information
  diverges logarithmically with the size of the data set we have
  observed}; the coefficient of this divergence counts the number of
parameters in the model, or more precisely the effective dimension of
the model class, and this provides a link to known results of Rissanen
and others.  We also can quantify the
complexity of processes that fall outside the conventional finite
dimensional models, and we show that these {\em more complex processes
  are characterized by a power--law rather than a logarithmic
  divergence of the predictive information}.

By analogy with the analysis of critical phenomena in statistical
physics, the separation of logarithmic from power--law divergences,
together with the measurement of coefficients and exponents for these
divergences, allows us to define ``universality classes'' for the
complexity of data streams. The power--law or nonparametric class of
processes may be crucial in real world learning tasks, where the
effective number of parameters becomes so large that asymptotic
results for finitely parameterizable models are inaccessible in
practice.  There is empirical evidence that simple physical systems
can generate dynamics in this complexity class, and there are hints
that language also may fall in this class.

Finally, we argue that {\em the divergent components of the predictive
  information provide a unique measure of complexity} that is
consistent with certain simple requirements.  This argument is in the
spirit of Shannon's original derivation of entropy as the unique
measure of available information. We believe that this uniqueness
argument provides a conclusive answer to the question of how one
should quantify the complexity of a process generating a time series.

With the evident cost of lengthening our discussion, we have tried to
give a self--contained presentation that develops our point of view,
uses simple examples to connect with known results, and then
generalizes and goes beyond these results.\footnote{Some of the basic
  ideas presented here, together with some connections to earlier
  work, can be found in brief preliminary reports (Bialek 1995; Bialek
  and Tishby 1999). The central results of the present work, however,
  were at best conjectures in these preliminary accounts.}  Even in
cases where at least the qualitative form of our results is known from
previous work, we believe that our point of view elucidates some
issues that may have been less the focus of earlier studies. Last but
not least, we explore the possibilities for connecting our theoretical
discussion with the experimental characterization of learning and
complexity in neural systems.

\section{A curious observation}

\begin{figure}[Ht]
  \centerline{\epsfxsize=0.8\hsize\epsffile{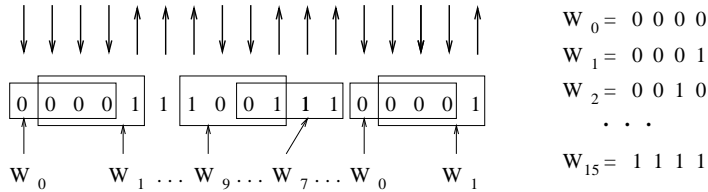}}
\caption[Calculating entropy of spin words.]{Calculating entropy of
  words of length $4$ in a chain of $17$ spins. For this chain,
  $n(W_0)=n(W_1)=n(W_3)=n(W_7)=n(W_{12})=n(W_{14})=2$,
  $n(W_8)=n(W_9)=1$, and all other frequencies are zero. Thus,
  $S(4)\approx2.95\; {\rm bits}$.}
\label{spins}
\end{figure}

Before starting the systematic analysis of the problem, we want to
motivate our discussion further by presenting results of some simple
numerical experiments.  Since most of the paper draws examples from
learning, here we consider examples from equilibrium statistical
mechanics. Suppose that we have a one dimensional chain of Ising spins
with the Hamiltonian given by
\begin{equation}
H=-\sum_{\rm i,j} J_{\rm ij} \sigma_{\rm i} \sigma_{\rm j},
\end{equation}
where the matrix of interactions $J_{\rm ij}$ is not restricted to
nearest neighbors---long range interactions are also allowed. One may
identify spins pointing upwards with $1$ and downwards with $0$, and
then a spin chain is equivalent to some sequence of binary digits.
This sequence consists of (overlapping) words of $N$ digits each,
$W_{\rm k}$, $k=0,1 \cdots 2^N-1$. There are $2^N$ such words total,
and they appear with very different frequencies $n(W_k)$ in the spin
chain [see Fig.~(\ref{spins}) for details].
If the number of spins is large, then counting these frequencies
provides a good empirical estimate of $P_N(W_k)$, the probability
distribution of different words of length $N$.  Then one can calculate
the entropy $S(N)$ of this probability distribution by the usual
formula
\begin{equation}
S(N)=-\sum_{k=0}^{2^N-1} P_N(W_k) \log_2 P_N(W_k)\;\;\;\;\;\;{\rm (bits)}.
\end{equation}
Note that this is not the entropy of a finite chain with length $N$;
instead it is the entropy of words or strings with length $N$ drawn
from a much longer chain.  Nonetheless, since entropy is an extensive
property, $S(N)$ is proportional asymptotically to $N$ for any spin
chain, that is $S(N ) \approx {\cal S}_0 \cdot N$.  The usual goal in
statistical mechanics is to understand this ``thermodynamic limit'' $N
\rightarrow \infty$, and hence to calculate the entropy density ${\cal
  S}_0$.  Different sets of interactions $J_{\rm ij}$ result in
different values of ${\cal S}_0$, but the qualitative result $S(N)
\propto N$ is true for all reasonable $\{J_{\rm ij}\}$.

\begin{figure}[Ht]
  \centerline{\epsfxsize=0.7\hsize\epsffile{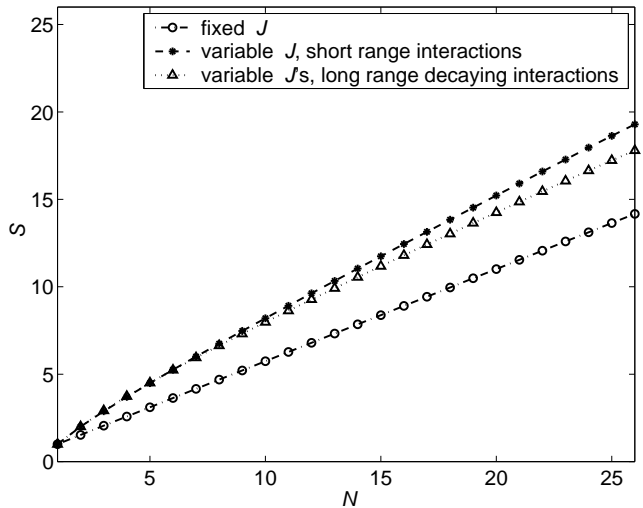}}
\caption[Entropy as a function of the word length]{Entropy as a
  function of the word length for spin chains with different
  interactions. Notice that all lines start from $S(N)=\log_2 2=1$
  since at the values of the coupling we investigated the correlation
  length is much smaller than the chain length ($1\cdot 10^9$ spins).}
\label{entr_lin}
\end{figure}

We investigated three different spin chains of one billion spins each.
As usual in statistical mechanics the probability of any configuration
of spins $\{\sigma_{\rm i}\}$ is given by the Boltzmann distribution,
\begin{equation}
P[\{\sigma_{\rm i}\}] \propto
\exp(-H/k_B T),
\end{equation}
where to normalize the scale of the $J_{\rm ij}$ we set $k_B T=1$.
For the first chain, only $J_{\rm i, i+1}=1$ was nonzero, and its
value was the same for all ${\rm i}$. The second chain was also
generated using the nearest neighbor interactions, but the value of
the coupling was reset every 400,000 spins by taking a random number
from a Gaussian distribution with zero mean and unit variance. In the
third case, we again reset the interactions at the same frequency, but
now interactions were long--ranged; the variances of coupling
constants decreased with the distance between the spins as $\langle
J_{\rm ij}^2 \rangle =1/({\rm i}-{\rm j})^2$. We plot $S(N)$ for all
these cases in Fig.~(\ref{entr_lin}),
and, of course, the asymptotically linear behavior is evident---the
extensive entropy shows no qualitative distinction among the three
cases we consider.

However, the situation changes drastically if we remove the asymptotic
linear contribution and focus on the {\em corrections} to extensive
behavior.  Specifically, we write $S(N) = {\cal S}_0 \cdot N +
S_1(N)$, and plot only the sublinear component $S_1(N)$ of the
entropy.
\begin{figure}[Ht]
  \centerline{\epsfxsize=0.7\hsize\epsffile{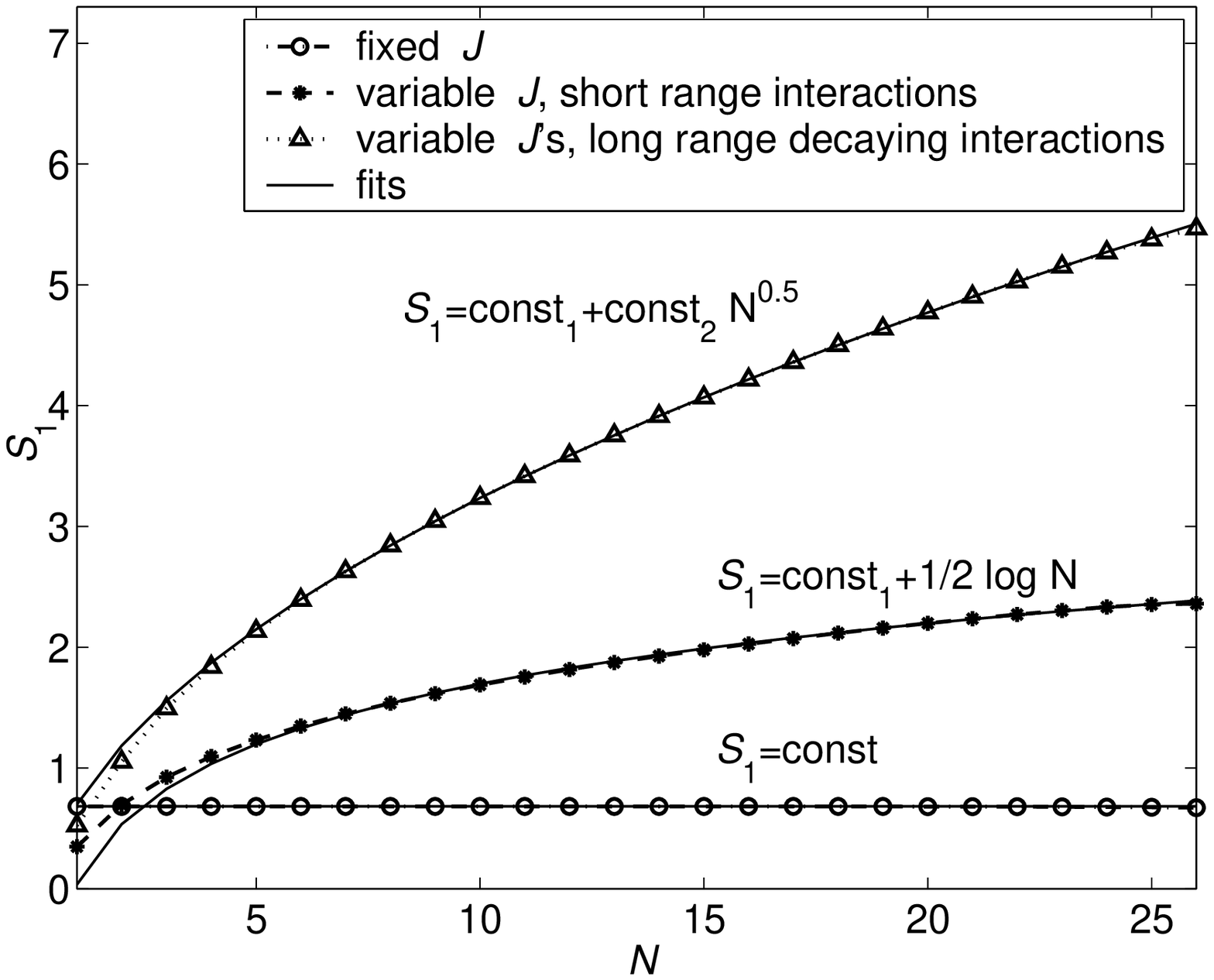}}
\caption{Subextensive part of the entropy as a function of the word
  length.}
\label{entr_subext}
\end{figure} 
As we see in Fig.~(\ref{entr_subext}), the three chains then exhibit
{\em qualitatively} different features: for the first one, $S_1$ is
constant; for the second one, it is logarithmic; and, for the third
one, it clearly shows a power--law behavior.

What is the significance of these observations? Of course, the
differences in the behavior of $S_1(N)$ must be related to the ways we
chose $J$'s for the simulations. In the first case, $J$ is fixed, and
if we see ${N}$ spins and try to predict the state of the ${ N+1}^{\rm
  st}$ spin, all that really matters is the state of the spin
$\sigma_{\rm N}$---there is nothing to ``learn'' from observations on
longer segments of the chain.  For the second chain, $J$ changes, and
the statistics of the spin--words are different in different parts of
the sequence. By looking at these statistics, one can ``learn'' the
coupling at the current position; this estimate improves the more
spins (longer words) we observe.  Finally, in the third case there are
many coupling constants that can be learned---as $N$ increases one
becomes sensitive to weaker correlations caused by interactions over
larger and larger distances.  So, intuitively, the qualitatively
different behaviors of $S_1(N)$ in the three plotted cases are
correlated with differences in the problem of learning the underlying
dynamics of the spin chains from observations on samples of the spins
themselves.  Much of this paper can be seen as expanding on and
quantifying this intuitive observation.

\section{Fundamentals}

The problem of prediction comes in various forms, as noted above.
Information theory allows us to treat the different notions of
prediction on the same footing.  The first step is to recognize that
all predictions are probabilistic---even if we can predict the
temperature at noon tomorrow, we should provide error bars or
confidence limits on our prediction. The next step is to remember
that, even before we look at the data, we know that certain futures
are more likely than others, and we can summarize this knowledge by a
prior probability distribution for the future.  Our observations on
the past lead us to a new, more tightly concentrated distribution, the
distribution of futures conditional on the past data. Different kinds
of predictions are different slices through or averages over this
conditional distribution, but information theory quantifies the
``concentration'' of the distribution without making any commitment as
to which averages will be most interesting.

Imagine that we observe a stream of data $x(t)$ over a time interval
$-T < t < 0$; let all of these past data be denoted by the shorthand
$x_{\rm past}$.  We are interested in saying something about the
future, so we want to know about the data $x(t)$ that will be observed
in the time interval $0 < t < T'$; let these future data be called
$x_{\rm future}$.  In the absence of any other knowledge, futures are
drawn from the probability distribution $P(x_{\rm future})$, while
observations of particular past data $x_{\rm past}$ tell us that
futures will be drawn from the conditional distribution $P(x_{\rm
  future} | x_{\rm past})$. The greater concentration of the
conditional distribution can be quantified by the fact that it has
smaller entropy than the prior distribution, and this reduction in
entropy is Shannon's definition of the information that the past
provides about the future.  We can write the average of this {\em
  predictive information} as
\begin{eqnarray}
{\cal I}_{\rm pred} (T,T') &=& 
{\Bigg\langle} \log_2 \left[ {{P(x_{\rm future}| x_{\rm past})} 
\over{P(x_{\rm future})}}\right]\Bigg\rangle
  \\ 
&=& -\langle\log_2 P(x_{\rm future})\rangle 
- \langle\log_2 P( x_{\rm past})\rangle
\nonumber\\
&&\,\,\,\,\,\,\,\,\,\, 
-\left[-\langle\log_2 P(x_{\rm future}, x_{\rm past})\rangle\right]\,,
\label{ents}
\end{eqnarray}
where $\langle \cdots \rangle$ denotes an average over the joint
distribution of the past and the future, $P(x_{\rm future} , x_{\rm
  past})$.

Each of the terms in Eq.~(\ref{ents}) is an entropy. Since we are
interested in predictability or generalization, which are associated
with some features of the signal persisting forever, we may assume
stationarity or invariance under time translations. Then the entropy
of the past data depends only on the duration of our observations, so
we can write $ -\langle\log_2 P( x_{\rm past})\rangle = S(T) $, and by
the same argument $-\langle\log_2 P( x_{\rm future})\rangle = S(T')$.
Finally, the entropy of the past and the future taken together is the
entropy of observations on a window of duration $T+T'$, so that $
-\langle\log_2 P(x_{\rm future} , x_{\rm past})\rangle = S(T+T')$.
Putting these equations together, we obtain
\begin{equation}
{\cal I}_{\rm pred}(T,T') = S(T) +S(T') - S(T+T') . \label{IpredandST}
\end{equation}

It is important to recall that mutual information is a symmetric
quantity.  Thus we can view ${\cal I}_{\rm pred}(T,T')$ either as the
information that a data segment of duration $T$ provides about the
future of length $T'$, {\em or} as the information that a data segment
of duration $T'$ provides about the immediate past of duration $T$.
This is a direct application of the definitions of information, but
seems counterintuitive: shouldn't it be more difficult to predict than
to postdict?  One can perhaps recover the correct intuition by
thinking about a large ensemble of question/answer pairs.  Prediction
corresponds to generating the answer to a given question, while
postdiction corresponds to generating the question that goes with a
given answer.  We know that guessing questions given answers is also a
hard problem,\footnote{This is the basis of the popular American
  television game show {\em Jeopardy!,} widely viewed as the most
  `intellectual' of its genre.} and can be just as challenging as the
more conventional problem of answering the questions themselves.  Our
focus here is on prediction because we want to make connections with
the phenomenon of generalization in learning, but it is clear that
generating a consistent interpretation of observed data may involve
elements of both prediction and postdiction [see, for example,
Eagleman and Sejnowski (2000)]; it is attractive that the information
theoretic formulation treats these problems symmetrically.

In the same way that the entropy of a gas at fixed density is
proportional to the volume, the entropy of a time series
(asymptotically) is proportional to its duration, so that
$\lim_{T\rightarrow\infty} {{S(T)}/ T} = {\cal S}_0$; entropy is an
extensive quantity.  But from Eq.~(\ref{IpredandST}) any extensive
component of the entropy cancels in the computation of the predictive
information: {\em predictability is a deviation from extensivity}.  If
we write $S(T) = {\cal S}_0 T +S_1(T)$, then Eq.~(\ref{IpredandST})
tells us that the predictive information is related {\em only} to the
nonextensive term $S_1(T)$.  Note that if we are observing a
deterministic system, then ${\cal S}_0 =0$, but this is independent of
questions about the structure of the subextensive term $S_1 (T)$. It
is attractive that information theory gives us a unified discussion of
prediction in deterministic and probabilistic cases.

We know two general facts about the behavior of $S_1(T)$.  First, the
corrections to extensive behavior are positive, $S_1(T) \geq 0$.
Second, the statement that entropy is extensive is the statement that
the limit
\begin{eqnarray}
\lim_{T\rightarrow\infty} {{S(T)}\over T} = {\cal S}_0 
\end{eqnarray}
exists, and for this to be true we must also have
\begin{eqnarray}
\lim_{T\rightarrow\infty} {{S_1(T)}\over T} = 0.
\end{eqnarray}
Thus the nonextensive terms in the entropy must be {\em sub}extensive,
that is they must grow with $T$ less rapidly than a linear function.
Taken together, these facts guarantee that the predictive information
is positive and subextensive.  Further, if we let the future extend
forward for a very long time, $T' \rightarrow \infty$, then we can
measure the information that our sample provides about the entire
future,
\begin{equation}
I_{\rm pred} (T) = \lim_{T' \rightarrow \infty} {\cal I}_{\rm
pred}(T,T')
= S_1 (T).
\end{equation}

Similarly, instead of
increasing the duration of the future to infinity we could have
considered the mutual information between a sample of length $T$ and
all of the infinite past. Then the {\em postdictive information}
also is equal to $S_1(T)$, and the symmetry between prediction and
postdiction is even more profound: not only is there symmetry between
questions and answers, but observations on a given period of time
provide the same amount of information about the historical path
that led to our observations as about the future that will unfold from them.
In some cases this statement becomes even stronger. For
example, if the subextensive entropy of a long discontinuous
observation of a total length $T$ with a gap of a duration $\delta T
\ll T$ is equal to $S_1(T)+ O({\delta T\over T})$, then the
subextensive entropy of the present is not only its information about
the past {\em or} the future, but also the information about the past
{\em and} the future. 

If we have been observing a time series for a (long) time $T$, then
the total amount of data we have collected in is measured by the
entropy $S(T)$, and at large $T$ this is given approximately by ${\cal
  S}_0 T$.  But the predictive information that we have gathered
cannot grow linearly with time, even if we are making predictions
about a future which stretches out to infinity. As a result, of the
total information we have taken in by observing $x_{\rm past}$, only a
vanishing fraction is of relevance to the prediction:
\begin{equation}
\lim_{T\rightarrow\infty} {{\rm Predictive\ Information} \over{\rm
Total\ Information}} = {{I_{\rm pred} (T)} \over {S(T)}}
\rightarrow 0. \label{chuck}
\end{equation}
In this precise sense, most of what we observe is irrelevant to the
problem of predicting the future.\footnote{We can think of
  Eq.~(\ref{chuck}) as a law of diminishing returns: although we
  collect data in proportion to our observation time $T$, a smaller
  and smaller fraction of this information is useful in the problem of
  prediction.  These diminishing returns are not due to a limited
  lifetime, since we calculate the predictive information assuming
  that we have a future extending forward to infinity.  A senior
  colleague points out that this is an argument for changing fields
  before becoming too expert.}

Consider the case where time is measured in discrete steps, so that we
have seen $N$ time points $x_1, x_2 , \cdots , x_N$. How much have we
learned about the underlying pattern in these data?  The more we know,
the more effectively we can predict the next data point $x_{N+1}$ and
hence the fewer bits we will need to describe the deviation of this
data point from our prediction: our accumulated knowledge about the
time series is measured by the degree to which we can compress the
description of new observations.  On average, the length of the code
word required to describe the point $x_{N+1}$, given that we have seen
the previous $N$ points, is given by
\begin{equation}
\ell(N) = -\langle \log_2 P(x_{N+1} | x_1 , x_2 , \cdots , x_N )
\rangle \,\,{\rm bits},
\label{condentropy}
\end{equation}
where the expectation value is taken over the joint distribution of
all the $N+1$ points, $P(x_1, x_2, \cdots , x_N, x_{N+1})$.  It is
easy to see that
\begin{eqnarray}
\ell(N) = S(N+1) - S(N) \approx 
{{\partial S (N)}\over {\partial N}} .
\end{eqnarray}
As we observe for longer times, we learn more and this word length
decreases.  It is natural to define a learning curve that measures
this improvement.  Usually we define learning curves by measuring the
frequency or costs of errors; here the cost is that our encoding of
the point $x_{N+1}$ is longer than it could be if we had perfect
knowledge.  This ideal encoding has a length which we can find by
imagining that we observe the time series for an infinitely long time,
$\ell_{\rm ideal} = \lim_{N\rightarrow\infty} \ell(N)$, but this is
just another way of defining the extensive component of the entropy
${\cal S}_0$. Thus we can define a learning curve
\begin{eqnarray}
\Lambda (N) &\equiv& \ell(N) - \ell_{\rm ideal} \\ 
&=& S(N+1) - S(N) - {\cal S}_0 \nonumber\\ 
&=& S_1(N+1) - S_1(N) \nonumber\\ 
&\approx&
{{\partial S_1 (N)}\over {\partial N}} 
= {{\partial I_{\rm pred}(N)}\over{\partial N}},
\label{unilcurve}
\end{eqnarray}
and we see once again that the extensive component of the entropy
cancels.

It is well known that the problems of prediction and compression are
related, and what we have done here is to illustrate one aspect of
this connection.  Specifically, if we ask how much one segment of a
time series can tell us about the future, the answer is contained in
the subextensive behavior of the entropy.  If we ask how much we are
learning about the structure of the time series, then the natural and
universally defined learning curve is related again to the
subextensive entropy: the learning curve is the derivative of the
predictive information.

This universal learning curve is connected to the more conventional
learning curves in specific contexts.  As an example (cf.~Section
\ref{testcase}), consider fitting a set of data points $\{ x_{\rm n} ,
y_{\rm n}\}$ with some class of functions $y = f(x; {\bgm \alpha}),$
where the $\bgm\alpha$ are unknown parameters that need to be learned;
we also allow for some Gaussian noise in our observation of the
$y_{\rm n}$.  Here the natural learning curve is the evolution of
$\chi^2$ for generalization as a function of the number of examples.
Within the approximations discussed below, it is straightforward to
show that as $N$ becomes large,
\begin{eqnarray}
\langle \chi^2(N) \rangle = {1\over {\sigma^2}} \langle
\left[ y - f(x;{\bgm \alpha})\right]^2\rangle \rightarrow 
(2 \ln2) \,\Lambda (N) + 1,
\end{eqnarray}
where $\sigma^2$ is the variance of the noise. Thus a more
conventional measure of performance at learning a function is equal to
the universal learning curve defined purely by information theoretic
criteria. In other words, if a learning curve is measured in the right
units, then its integral represents the amount of the useful
information accumulated. Then the subextensivity of $S_1$ guarantees
that the learning curve decreases to zero as $N \to \infty$.

Different quantities related to the subextensive entropy have been
discussed in several contexts. For example, the code length $\ell(N)$
has been defined as a learning curve in the specific case of neural
networks (Opper and Haussler 1995) and has been termed the
``thermodynamic dive'' (Crutchfield and Shalizi 1998) and ``$N^{\rm
  th}$ order block entropy'' (Grassberger 1986). The universal
learning curve $\Lambda(N)$ has been studied as the ``expected
instantaneous information gain'' by Haussler et al.~(1994). Mutual
information between all of the past and all of the future (both
semi--infinite) is known also as the ``excess entropy,'' ``effective
measure complexity,'' ``stored information,'' and so on [see Shalizi
and Crutchfield (1999) and references therein, as well as the
discussion below]. If the data allow a description by a model with
finite (and in some cases also infinite) number of parameters, then
mutual information between the data and the parameters is of interest.
This is easily shown to be equal to the predictive information about
all of the future, and it is also the ``cumulative information gain''
(Haussler et al.~1994) or the ``cumulative relative entropy risk''
(Haussler and Opper 1997).  Investigation of this problem can be
traced back to Renyi (1964) and Ibragimov and Hasminskii (1972), and
some particular topics are still being discussed (Haussler and Opper
1995, Opper and Haussler 1995, Herschkowitz and Nadal 1999). In
certain limits, decoding a signal from a population of $N$ neurons can
be thought of as `learning' a parameter from $N$ observations, with a
parallel notion of information transmission (Brunel and Nadal 1998,
Kang and Sompolinsky 2001).  In addition, as noted already, the
subextensive component of the description length (Rissanen 1978, 1989,
1996, Clarke and Barron 1990) averaged over a class of allowed models
also is similar to the predictive information.  What is important is
that {\em the predictive information or subextensive entropy is
  related to all these quantities}, and that {\em it can be defined
  for any process without a reference to a class of models}.  It is
this universality that we find appealing, and this universality is
strongest if we focus on the limit of long observation times.
Qualitatively, in this regime ($T\rightarrow\infty$) we expect the
predictive information to behave in one of three different ways, as
illustrated by the Ising models above: it may either stay finite, or
grow to infinity together with $T$; in the latter case the rate of
growth may be slow (logarithmic) or fast (sublinear power) [see Barron
and Cover (1991) for a similar classification in the framework of the
Minimal Description Length (MDL) analysis].

The first possibility, $\lim_{T\rightarrow\infty} I_{\rm pred} (T) = $
constant, means that no matter how long we observe we gain only a
finite amount of information about the future. This situation
prevails, for example, when the dynamics are too regular: for a purely
periodic system, complete prediction is possible once we know the
phase, and if we sample the data at discrete times this is a finite
amount of information; longer period orbits intuitively are more
complex and also have larger $I_{\rm pred}$, but this doesn't change
the limiting behavior $\lim_{T\rightarrow\infty} I_{\rm pred} (T) =$
constant.

Alternatively, the predictive information can be small when the
dynamics are irregular but the best predictions are controlled only by
the immediate past, so that the correlation times of the observable
data are finite [see, for example, Crutchfield and Feldman (1997) and
the fixed $J$ case in Fig.~(\ref{entr_subext})].  Imagine, for
example, that we observe $x(t)$ at a series of discrete times
$\{t_{\rm n}\}$, and that at each time point we find the value $x_{\rm
  n}$. Then we always can write the joint distribution of the $N$ data
points as a product,
\begin{eqnarray}
P(x_1 , x_2 , \cdots , x_N ) &=& P(x_1 ) P(x_2 | x_1) 
P(x_3 | x_2 , x_1) \cdots . 
\end{eqnarray}
For Markov processes, what we observe at $t_{\rm n}$ depends only on
events at the previous time step $t_{\rm n-1}$, so that
\begin{eqnarray}
P(x_{\rm n} | \{x_{\rm 1\leq i \leq n-1}\}) &=& 
P(x_{\rm n} | x_{\rm n-1}) ,
\end{eqnarray}
and hence the predictive information reduces to
\begin{eqnarray}
I_{\rm pred} = \Bigg\langle \log_2 \left[ {{P(x_{\rm n}|x_{\rm n-1})}
\over{P(x_{\rm n})}} \right] \Bigg\rangle .
\end{eqnarray}
The maximum possible predictive information in this case is the
entropy of the distribution of states at one time step, which in turn
is bounded by the logarithm of the number of accessible states. To
approach this bound the system must maintain memory for a long time,
since the predictive information is reduced by the entropy of the
transition probabilities. Thus systems with more states and longer
memories have larger values of $I_{\rm pred}$.

More interesting are those cases in which $I_{\rm pred}(T)$ diverges
at large $T$. In physical systems we know that there are critical
points where correlation times become infinite, so that optimal
predictions will be influenced by events in the arbitrarily distant
past. Under these conditions the predictive information can grow
without bound as $T$ becomes large; for many systems the divergence is
logarithmic, $I_{\rm pred} (T\rightarrow\infty) \propto \ln T$, as for
the variable $J_{\rm ij}$, short range Ising model of
Figs.~(\ref{entr_lin}, \ref{entr_subext}). Long range correlation also
are important in a time series where we can learn some underlying
rules.  It will turn out that when the set of possible rules can be
described by a finite number of parameters, the predictive information
again diverges logarithmically, and the coefficient of this divergence
counts the number of parameters.  Finally, a faster growth is also
possible, so that $I_{\rm pred} (T\rightarrow\infty) \propto
T^\alpha$, as for the variable $J_{\rm ij}$ long range Ising model,
and we shall see that this behavior emerges from, for example,
nonparametric learning problems.

\section{Learning and predictability}

Learning is of interest precisely in those situations where
correlations or associations persist over long periods of time. In the
usual theoretical models, there is some rule underlying the observable
data, and this rule is valid forever; examples seen at one time inform
us about the rule, and this information can be used to make
predictions or generalizations. The predictive information quantifies
the average generalization power of examples, and we shall see that
there is a direct connection between the predictive information and
the complexity of the possible underlying rules.

\subsection{A test case}
\label{testcase}

Let us begin with a simple example already mentioned above.  We
observe two streams of data $x$ and $y$, or equivalently a stream of
pairs $(x_{\rm 1} , y_{\rm 1})$, $(x_{\rm 2} , y_{\rm 2})$, $\cdots$ ,
$(x_{\rm N} , y_{\rm N})$.  Assume that we know in advance that the
$x$'s are drawn independently and at random from a distribution
$P(x)$, while the $y$'s are noisy versions of some function acting on
$x$,
\begin{eqnarray}
y_{\rm n} = f(x_{\rm n} ; {\bgm\alpha} ) + \eta_{\rm n} ,
\end{eqnarray}
where $f(x; {\bgm\alpha})$ is a class of functions parameterized by
$\bgm\alpha$, and $\eta_{\rm n}$ is noise, which for simplicity we
will assume is Gaussian with known standard deviation $\sigma$.  We
can even start with a {\em very} simple case, where the function class
is just a linear combination of basis functions, so that
\begin{eqnarray}
f(x; {\bgm\alpha}) = \sum_{\rm \mu =1}^K \alpha_\mu \phi_\mu (x) .
\end{eqnarray}
The usual problem is to estimate, from $N$ pairs $\{x_{\rm i} , y_{\rm
  i}\}$, the values of the parameters $\bgm\alpha$; in favorable cases
such as this we might even be able to find an effective regression
formula.  We are interested in evaluating the predictive information,
which means that we need to know the entropy $S(N)$.  We go through
the calculation in some detail because it provides a model for the
more general case.

To evaluate the entropy $S(N)$ we first construct the probability
distribution $P(x_{\rm 1}, y_{\rm 1}, x_{\rm 2}, y_{\rm 2}, \cdots ,
x_{\rm N} , y_{\rm N })$. The same set of rules applies to the whole
data stream, which here means that the same parameters $\bgm\alpha$
apply for all pairs $\{x_{\rm i}, y_{\rm i}\}$, but these parameters
are chosen at random from a distribution ${\cal P}({\bgm\alpha})$ at
the start of the stream. Thus we write
\begin{eqnarray}
&&P(x_{\rm 1}, y_{\rm 1}, x_{\rm 2} , y_{\rm 2}, \cdots , x_{\rm
N} , y_{\rm N }) 
\nonumber\\
&&\,\,\,\,\,= \int d^K \alpha \,
P(x_{\rm 1}, y_{\rm 1}, x_{\rm 2} , y_{\rm 2}, \cdots , x_{\rm N}
, y_{\rm N }  | {\bgm\alpha}) {\cal P}({\bgm\alpha}) \,,
\label{start} 
\end{eqnarray}
and now we need to construct the conditional distributions for fixed
$\bgm\alpha$. By hypothesis each $x$ is chosen independently, and once
we fix $\bgm\alpha$ each $y_{\rm i}$ is correlated only with the
corresponding $x_{\rm i}$, so that we have
\begin{eqnarray}
P(x_{\rm 1}, y_{\rm 1}, x_{\rm 2} , y_{\rm 2}, \cdots , x_{\rm N}
, y_{\rm N } | {\bgm\alpha}) = \prod_{\rm i =1}^{\rm N} \left[
P(x_{\rm i})\, P(y_{\rm i} | x_{\rm i} ; {\bgm\alpha})\right]\,.
\end{eqnarray}
Further, with the simple assumptions above about the class of
functions and Gaussian noise, the conditional distribution of $y_{\rm
  i}$ has the form
\begin{eqnarray}
P(y_{\rm i} | x_{\rm i} ; {\bgm\alpha}) = 
{1\over\sqrt{2 \pi\sigma^2}} 
\exp\left[ -{1\over{2\sigma^2}} \left( y_{\rm i} - \sum_{\mu =1}^K 
\alpha_\mu \phi_\mu(x_{\rm i})\right)^2 \right] \,.
\end{eqnarray}
Putting all these factors together,
\begin{eqnarray}
&&P(x_{\rm 1}, y_{\rm 1}, x_{\rm 2} , y_{\rm 2}, \cdots , x_{\rm
N} , y_{\rm N }) 
\nonumber\\ 
&&\,\,\,\,\,= \left[ \prod_{\rm i =1}^{\rm N} P(x_{\rm i}) \right] 
\left( {1\over \sqrt{2\pi\sigma^2}}\right)^{N} \int d^K \alpha \, 
{\cal P}({\bgm\alpha})
\exp\left[
-{1\over{2\sigma^2}} \sum_{\rm i=1}^{\rm N} y_{\rm i}^2\right] 
\nonumber\\
&&\,\,\,\,\,\,\,\,\,\,\times 
\exp\left[
-{N\over 2} \sum_{\mu , \nu =1}^K A_{\mu\nu} (\{x_{\rm i}\})
\alpha_\mu\alpha_\nu 
+ N \sum_{\mu=1}^K B_\mu(\{x_{\rm i} , y_{\rm i}\}) 
\alpha_\mu \right] ,
\end{eqnarray}
where
\begin{eqnarray}
A_{\mu\nu} (\{x_{\rm i}\}) &=& {1\over{\sigma^2 N}}
\sum_{\rm i=1}^{\rm N} \phi_\mu(x_{\rm i} ) \phi_\nu (x_{\rm i}) ,
\,\,\,\,{\rm and}
\\
B_\mu(\{x_{\rm i} , y_{\rm i}\}) &=& {1\over{\sigma^2 N}}\sum_{\rm
i=1}^{\rm N} y_{\rm i} \phi_\mu(x_{\rm i} ) .
\end{eqnarray}
Our placement of the factors of $N$ means that both $A_{\mu\nu}$ and
$B_\mu$ are of order unity as $N\rightarrow\infty$.  These quantities
are empirical averages over the samples $\{ x_{\rm i} , y_{\rm i}\}$,
and if the $\phi_\mu$ are well behaved we expect that these empirical
means converge to expectation values for most realizations of the
series $\{x_{\rm i}\}$:
\begin{eqnarray}
\lim_{N\rightarrow\infty} A_{\mu\nu} (\{x_{\rm i}\})
 &=&A_{\mu\nu}^\infty = {1\over{\sigma^2 }}\int dx P(x) \phi_\mu(x  )
\phi_\nu (x )\,,
\\
\lim_{N\rightarrow\infty} B_\mu(\{x_{\rm i} , y_{\rm i}\})
&=&B_\mu^\infty  = \sum_{\nu =1}^K A_{\mu\nu}^\infty
\bar{\alpha}_\nu\, ,
\end{eqnarray}
where $\bar{\bgm\alpha}$ are the parameters that actually gave rise to
the data stream $\{ x_{\rm i} , y_{\rm i}\}$.  In fact we can make the
same argument about the terms in $\sum y_{\rm i}^2$,
\begin{eqnarray}
\lim_{N\rightarrow\infty} \sum_{\rm i=1}^{\rm N} y_{\rm i}^2 = N
\sigma^2 \left[ \sum_{\mu,\nu =1}^K {\bar\alpha}_\mu
A_{\mu\nu}^\infty {\bar\alpha}_\nu + 1 \right].
\end{eqnarray}
Conditions for this convergence of empirical means to expectation
values are at the heart of learning theory.  Our approach here is
first to assume that this convergence works, then to examine the
consequences for the predictive information, and finally to address
the conditions for and implications of this convergence breaking down.

Putting the different factors together, we obtain
\begin{eqnarray}
&&P(x_{\rm 1}, y_{\rm 1}, x_{\rm 2} , y_{\rm 2}, \cdots , x_{\rm
N} , y_{\rm N }) 
\nonumber
\\ 
&&\,\,\,\,\,{\widetilde\rightarrow} \left[ \prod_{\rm i =1}^{\rm N} 
P(x_{\rm i}) \right] \left( {1\over \sqrt{2\pi\sigma^2}}\right)^{N} 
\int d^K \alpha {\cal P}({\bgm\alpha}) 
\exp\left[
-N E_N({\bgm\alpha} ; \{x_{\rm i} , y_{\rm i}\}) \right] ,
\label{integral2}
\nonumber\\
&&
\end{eqnarray}
where the effective ``energy'' per sample is given by
\begin{eqnarray}
E_N({\bgm\alpha} ; \{x_{\rm i} , y_{\rm i}\}) =
{1\over 2}  + 
{1\over 2} \sum_{\mu , \nu =1}^K (\alpha_\mu - {\bar\alpha}_\mu )
A_{\mu\nu}^\infty (\alpha_\nu - {\bar\alpha}_\nu ) .
\end{eqnarray}
Here we use the symbol ${\widetilde\to}$ to indicate that we not only
take the limit of large $N$, but also neglect the fluctuations. Note
that in this approximation the dependence on the sample points
themselves is hidden in the definition of $\bar{\bgm\alpha}$ as being
the parameters that generated the samples.

The integral that we need to do in Eq.~(\ref{integral2}) involves an
exponential with a large factor $N$ in the exponent; the energy $E_N$
is of order unity as $N\rightarrow\infty$.  This suggests that we
evaluate the integral by a saddle point or steepest descent
approximation [similar analyses were performed by Clarke and Barron
(1990), by MacKay (1992), and by Balasubramanian (1997)]:
\begin{eqnarray}
&&
\int d^K \alpha {\cal P}({\bgm\alpha})\exp
\left[-N E_N({\bgm\alpha} ; \{x_{\rm i}, y_{\rm i}\}) \right] 
\nonumber 
\approx 
{\cal P}({\bgm\alpha}_{\rm cl})
\\
&&\,\,\,\,
\times \exp \left[-NE_N({\bgm\alpha}_{\rm cl} ; 
\{x_{\rm i} , y_{\rm i}\})
-{K\over 2} \ln {N\over {2\pi}}  - 
{1\over2} \ln\det {\cal F}_N + \cdots \right],
\end{eqnarray}
where ${\bgm\alpha}_{\rm cl}$ is the ``classical'' value of
${\bgm\alpha}$ determined by the extremal conditions
\begin{eqnarray}
{{\partial E_N({\bgm\alpha} ; \{x_{\rm i} , y_{\rm i}\})}
\over{\partial\alpha_\mu}}\Bigg|_{{\bgm\alpha} ={\bgm\alpha}_{\rm
cl}} = 0 ,
\end{eqnarray}
the matrix ${\cal F}_N$ consists of the second derivatives of $E_N$,
\begin{eqnarray}
{\cal F}_N = 
{{\partial^2 E_N({\bgm\alpha} ; \{x_{\rm i} , y_{\rm i}\})}
\over{\partial\alpha_\mu\partial\alpha_\nu}}\Bigg|_{{\bgm\alpha}
={\bgm\alpha}_{\rm cl}} ,
\end{eqnarray}
and $\cdots$ denotes terms that vanish as $N\rightarrow\infty$.  If we
formulate the problem of estimating the parameters $\bgm\alpha$ from
the samples $\{x_{\rm i}, y_{\rm i}\}$, then as $N\rightarrow\infty$
the matrix $N{\cal F}_N$ is the Fisher information matrix (Cover and
Thomas 1991); the eigenvectors of this matrix give the principal axes
for the error ellipsoid in parameter space, and the (inverse)
eigenvalues give the variances of parameter estimates along each of
these directions. The classical ${\bgm\alpha}_{\rm cl}$ differs from
$\bar{\bgm\alpha}$ only in terms of order $1/N$; we neglect this
difference and further simplify the calculation of leading terms as
$N$ becomes large.  After a little more algebra, then, we find the
probability distribution we have been looking for:
\begin{eqnarray}
&&P(x_{\rm 1}, y_{\rm 1}, x_{\rm 2} , y_{\rm 2}, \cdots , x_{\rm N} , 
y_{\rm N }) 
\nonumber
\\ 
&&\,\,\,\,\,\widetilde\rightarrow 
\left[ \prod_{\rm i =1}^{\rm N} P(x_{\rm i}) \right] 
{1\over {Z_A}} {\cal P}(\bar{\bgm\alpha}) \exp \left[ -{N\over 2} 
\ln(2\pi {\rm e}\sigma^2) - {K\over 2} \ln N + \cdots \right],
\label{finalp}
\end{eqnarray}
where the normalization constant
\begin{eqnarray}
{Z_A} = \sqrt{(2\pi)^K\det A^\infty}.
\end{eqnarray}
Again we note that the sample points $\{ x_{\rm i}, y_{\rm i} \}$ are
hidden in the value of $\bar{\bgm\alpha}$ that gave rise to these
points.\footnote{We emphasize once more that there are two
  approximations leading to Eq.~(\ref{finalp}).  First, we have
  replaced empirical means by expectation values, neglecting
  fluctuations associated with the particular set of sample points $\{
  x_{\rm i} , y_{\rm i}\}$.  Second, we have evaluated the average
  over parameters in a saddle point approximation.  At least under
  some conditions, both of these approximations become increasingly
  accurate as $N\rightarrow \infty$, so that this approach should
  yield the asymptotic behavior of the distribution and hence the
  subextensive entropy at large $N$.  Although we give a more detailed
  analysis below, it is worth noting here how things can go wrong. The
  two approximations are independent, and we could imagine that
  fluctuations are important but saddle point integration still works,
  for example. Controlling the fluctuations turns out to be exactly
  the question of whether our finite parameterization captures the
  true dimensionality of the class of models, as discussed in the
  classic work of Vapnik, Chervonenkis, and others [see Vapnik (1998)
  for a review].  The saddle point approximation can break down
  because the saddle point becomes unstable or because multiple saddle
  points become important.  It will turn out that instability is
  exponentially improbable as $N\rightarrow\infty$, while multiple
  saddle points are a real problem in certain classes of models, again
  when counting parameters doesn't really measure the complexity of
  the model class.}

To evaluate the entropy $S(N)$ we need to compute the expectation
value of the (negative) logarithm of the probability distribution in
Eq.~(\ref{finalp}); there are three terms.  One is constant, so
averaging is trivial.  The second term depends only on the $x_{\rm
  i}$, and because these are chosen independently from the
distribution $P(x)$ the average again is easy to evaluate.  The third
term involves $\bar{\bgm\alpha}$, and we need to average this over the
joint distribution $P(x_{\rm 1}, y_{\rm 1}, x_{\rm 2} , y_{\rm 2},
\cdots , x_{\rm N} , y_{\rm N })$. As above, we can evaluate this
average in steps: first we choose a value of the parameters
$\bar{\bgm\alpha}$, then we average over the samples given these
parameters, and finally we average over parameters.  But because
$\bar{\bgm\alpha}$ is defined as the parameters that generate the
samples, this stepwise procedure simplifies enormously. The end result
is that
\begin{equation}
S(N) = N\left[ S_x +{1\over 2} \log_2 (2\pi{\rm e}\sigma^2)\right]
+ {K\over 2} \log_2 N + S_{\bgms\alpha} + \langle \log_2 {Z_A}
\rangle_{\bgms \alpha} + \cdots ,
\label{Stoy}
\end{equation}
where $\langle\cdots\rangle_{\bgms \alpha}$ means averaging over
parameters, $S_x$ is the entropy of the distribution of $x$,
\begin{eqnarray}
S_x = -\int dx \,P(x) \log_2 P(x) ,
\end{eqnarray}
and similarly for the entropy of the distribution of parameters,
\begin{eqnarray}
S_{\bgms\alpha} = -\int d^K \alpha \, {\cal P}({\bgm\alpha}) \log_2
{\cal P}({\bgm\alpha}) .
\end{eqnarray}

The different terms in the entropy Eq.~(\ref{Stoy}) have a
straightforward interpretation. First we see that the extensive term
in the entropy,
\begin{eqnarray}
{\cal S}_0 = S_x +{1\over 2} \log_2 (2\pi{\rm e}\sigma^2) ,
\label{extens1}
\end{eqnarray}
reflects contributions from the random choice of $x$ and from the
Gaussian noise in $y$; these extensive terms are independent of the
variations in parameters $\bgm\alpha$, and these would be the only
terms if the parameters were not varying (that is, if there were
nothing to learn).  There also is a term which reflects the entropy of
variations in the parameters themselves, $S_{\bgms\alpha}$.  This
entropy is not invariant with respect to coordinate transformations in
the parameter space, but the term $\langle\log_2
Z_A\rangle_{\bgms\alpha}$ compensates for this noninvariance.
Finally, and most interestingly for our purposes, the subextensive
piece of the entropy is dominated by a logarithmic divergence,
\begin{equation}\label{S1example}
S_1(N) \rightarrow {K\over 2}\log_2 N \;\;\;{\rm (bits)}.
\end{equation}
The coefficient of this divergence counts the number of parameters
independent of the coordinate system that we choose in the parameter
space. Furthermore, this result does not depend on the set of basis
functions $\{\phi_{\mu}(x)\}$. This is a hint that the result in
Eq.~(\ref{S1example}) is more universal than our simple example.

\subsection{Learning a parameterized distribution}
\label{learn_distr_sec}

The problem discussed above is an example of supervised learning: we
are given examples of how the points $x_{\rm n}$ map into $y_{\rm n}$,
and from these examples we are to induce the association or functional
relation between $x$ and $y$.  An alternative view is that pair of
points $(x,y)$ should be viewed as a vector $\vec x$, and what we are
learning is the distribution of this vector.  The problem of learning
a distribution usually is called unsupervised learning, but in this
case supervised learning formally is a special case of unsupervised
learning; if we admit that all the functional relations or
associations that we are trying to learn have an element of noise or
stochasticity, then this connection between supervised and
unsupervised problems is quite general.

Suppose a series of random vector variables $\{{\vec x}_{\rm i}\}$ are
drawn independently from the same probability distribution $Q({\vec x}
| {\bgm\alpha})$, and this distribution depends on a (potentially
infinite dimensional) vector of parameters $\bgm{\alpha}$.  As above,
the parameters are unknown, and before the series starts they are
chosen randomly from a distribution ${\cal P} (\bgm{\alpha})$. With no
constraints on the densities ${\cal P} (\bgm{\alpha})$ or $Q({\vec x}
| {\bgm\alpha})$ it is impossible to derive any regression formulas
for parameter estimation, but one can still say something about the
entropy of the data series and thus the predictive information. For a
finite dimensional vector of parameters $\bgm{\alpha}$ the literature
on bounding similar quantities is especially rich (Haussler et al.\ 
1994, Wong and Shen 1995, Haussler and Opper 1995, Haussler and Opper
1997 and references therein), and related asymptotic calculations have
been done (Clarke and Barron 1990, MacKay 1992, Balasubramanian 1997).

We begin with the definition of entropy
\begin{eqnarray}
S(N)&\equiv& S[\left\{ \vec{x}_{\rm i}\right\}]=
\nonumber
\\
&&-\int d{\vec x}_{\rm 1}\cdots d{\vec x}_{\rm N} \,
P({\vec x}_{\rm 1} , {\vec x}_{\rm 2} , \cdots , {\vec x}_{\rm N}) 
\,\log_2 P({\vec x}_{\rm 1} , {\vec x}_{\rm 2} , \cdots , {\vec x}_{\rm N}).
\end{eqnarray}
By analogy with Eq.~(\ref{start}) we then write
\begin{eqnarray}
P({\vec x}_{\rm 1} , {\vec x}_{\rm 2} , \cdots , {\vec x}_{\rm N})
= \int d^K\alpha {\cal P} ({\bgm\alpha}) \prod_{\rm i=1}^{\rm N} 
Q({\vec x}_{\rm i} | {\bgm\alpha}) .
\end{eqnarray}
Next, combining the last two equations and rearranging the order of
integration, we can rewrite $S(N)$ as
\begin{eqnarray}
S(N) &=& 
-\int d^K\bar{\bgm\alpha} {\cal P} (\bar{\bgm\alpha}) \left\{
\int d{\vec x}_{\rm 1}\cdots d{\vec x}_{\rm N} \,
\prod_{\rm j=1}^{\rm N} Q({\vec x}_{\rm j} | \bar{\bgm\alpha})
\;\log_2 P(\{{\vec x}_{\rm i}\})
\right\}.\nonumber\\
&&
\label{Srewritten}
\end{eqnarray}

Equation (\ref{Srewritten}) allows an easy interpretation. There is
the `true' set of parameters $\bar{\bgm\alpha}$ that gave rise to the
data sequence ${\vec x}_{\rm 1}\cdots {\vec x}_{\rm N}$ with the
probability $\prod_{\rm j=1}^{\rm N} Q({\vec x}_{\rm j} |
\bar{\bgm\alpha})$. We need to average $\log_2 P({\vec x}_{\rm
  1}\cdots {\vec x}_{\rm N})$ first over all possible realizations of
the data keeping the true parameters fixed, and then over the
parameters $\bar{\bgm\alpha}$ themselves. With this interpretation in
mind, the joint probability density, the logarithm of which is being
averaged, can be rewritten in the following useful way:
\begin{eqnarray}
P({\vec x}_{\rm 1} , \cdots , {\vec x}_{\rm N})
&=& \prod_{\rm j=1}^{\rm N} Q({\vec x}_{\rm j} | \bar{\bgm\alpha})
\int d^K\alpha {\cal P} ({\bgm\alpha}) \prod_{\rm i=1}^{\rm N} \left[
{{Q({\vec x}_{\rm i} | {\bgm\alpha})}\over {Q({\vec x}_{\rm i} |
\bar{\bgm\alpha})}} \right] 
\nonumber
\\ 
&=& \prod_{\rm j=1}^{\rm N}
Q({\vec x}_{\rm j} | \bar{\bgm\alpha}) 
\int d^K\alpha {\cal P} ({\bgm\alpha})\exp\left[
-N{\cal E}_N({\bgm\alpha} ; \{{\vec x}_{\rm i}\} )\right],
\label{pfn}
\\
{\cal E}_N({\bgm\alpha} ; \{{\vec x}_{\rm i}\} ) 
&=&  - {1\over N} \sum_{\rm i=1}^{\rm N} \ln\left[
{{Q({\vec x}_{\rm i} | {\bgm\alpha})}\over {Q({\vec x}_{\rm i} |
\bar{\bgm\alpha})}} \right] .
\label{empavg}
\end{eqnarray}
Since, by our interpretation, $\bar{\bgm\alpha}$ are the true
parameters that gave rise to the particular data $\{{\vec x}_{\rm
  i}\}$, we may expect that the empirical average in
Eq.~(\ref{empavg}) will converge to the corresponding expectation
value, so that
\begin{equation}\label{fn}
{\cal E}_N({\bgm\alpha} ; \{{\vec x}_{\rm i}\} ) = 
 - \int d^Dx Q(x|\bar{\bgm\alpha})  \ln\left[
{{Q({\vec x} | {\bgm\alpha})}\over {Q({\vec x} |
\bar{\bgm\alpha})}} \right] 
-\psi ({\bgm\alpha},\bar{\bgm\alpha} ; \{ x_{\rm i} \} ),
\end{equation}
where $\psi \rightarrow 0$ as $N\rightarrow \infty$; here we neglect
$\psi$, and return to this term below.

The first term on the right hand side of Eq.~(\ref{fn}) is the
Kullback--Leibler divergence, $D_{\rm KL}
({\bgm{\bar\alpha}}||{\bgm\alpha})$, between the true distribution
characterized by parameters $\bar{\bgm\alpha}$ and the possible
distribution characterized by $\bgm\alpha$.  Thus at large $N$ we have
\begin{eqnarray}
P({\vec x}_{\rm 1} , {\vec x}_{\rm 2} , \cdots , {\vec x}_{\rm N})
\widetilde\rightarrow \prod_{\rm j=1}^{\rm N} Q({\vec x}_{\rm j} |
\bar{\bgm\alpha}) \int d^K\alpha {\cal P} ({\bgm\alpha})
\exp\left[-ND_{\rm KL} (\bar{\bgm\alpha}||{\bgm\alpha}) \right] ,
\end{eqnarray}
where again the notation $\widetilde\rightarrow$ reminds us that we
are not only taking the limit of large $N$ but also making another
approximation in neglecting fluctuations.  By the same arguments as
above we can proceed (formally) to compute the entropy of this
distribution, and we find
\begin{eqnarray}
S(N) &\approx& {\cal S}_0 \cdot N + S^{({\rm a})}_1(N),
\\ 
{\cal S}_0 &=& \int d^K\alpha {\cal P}({\bgm\alpha}) \left[
-\int d^Dx Q({\vec x}|{\bgm\alpha}) \log_2 Q({\vec
x}|{\bgm\alpha}) \right], \,\,\,\,{\rm and}
\label{S0}
\\ 
S^{({\rm a})}_1(N)
&=& -\int d^K {\bar\alpha} {\cal P}(\bar{\bgm\alpha}) \log_2 \left[ \int
d^K\alpha P({\bgm\alpha}) {\rm e}^{-ND_{\rm KL}
(\bar{\bgms\alpha}||{\bgms\alpha})} \right] . 
\label{S1annealed}
\end{eqnarray}
Here $S^{({\rm a})}_1$ is an approximation to $S_1$ that neglects
fluctuations $\psi$. This is the same as the annealed approximation in
the statistical mechanics of disordered systems, as has been used
widely in the study of supervised learning problems (Seung et al.\ 
1992). Thus we can identify the particular data sequence ${\vec
  x}_{\rm 1}\cdots{\vec x}_{\rm N}$ with the disorder, ${\cal
  E}_N({\bgm\alpha} ; \{{\vec x}_{\rm i}\} )$ with the energy of the
quenched system, and $D_{\rm KL} (\bar{\bgm\alpha}||{\bgm\alpha})$
with its annealed analogue.

The extensive term ${\cal S}_0$, Eq.~(\ref{S0}), is the average
entropy of a distribution in our family of possible distributions,
generalizing the result of Eq.~(\ref{extens1}). The subextensive terms
in the entropy are controlled by the $N$ dependence of the partition
function
\begin{equation}\label{Z}
Z ({\bar{\bgm\alpha}} ;N) = \int d^K\alpha {\cal P} ({\bgm\alpha})
\exp\left[ -ND_{\rm KL} ( \bar{\bgm\alpha} || {\bgm\alpha}
)\right] ,
\end{equation}
and $S_1(N) = -\langle\log_2 Z ({\bar{\bgm\alpha}}
;N)\rangle_{\bar{\bgms\alpha}}$ is analogous to the free energy. Since
what is important in this integral is the Kullback--Leibler (KL)
divergence between different distributions, it is natural to ask about
the density of models that are KL divergence $D$ away from the target
$\bar{\bgm\alpha}$,
\begin{eqnarray}
\rho (D;{\bar{\bgm\alpha}}) = \int d^K\alpha  {\cal P} ({\bgm\alpha})
\delta [D - D_{\rm KL} ( \bar{\bgm\alpha} || {\bgm\alpha} )];
\end{eqnarray}
this density could be very different for different
targets.\footnote{If parameter space is compact, then a related
  description of the space of targets based on metric entropy, also
  called Kolmogorov's $\epsilon$-entropy, is used in the
  literature [see, for example, Haussler and Opper (1997)]. Metric
  entropy is the logarithm of the smallest number of disjoint parts of
  the size not greater than $\epsilon$ into which the target space can
  be partitioned.} The density of divergences is normalized because
the original distribution over parameter space, $P({\bgm\alpha})$, is
normalized,
\begin{eqnarray}
\int dD \rho (D;{\bar{\bgm\alpha}}) = \int d^K\alpha
{\cal P} ({\bgm\alpha}) = 1.
\label{normalize}
\end{eqnarray}
Finally, the partition function takes the simple form
\begin{equation}\label{ZD}
Z ({\bar{\bgm\alpha}} ;N) = \int dD \rho (D;{\bar{\bgm\alpha}})
\exp[-ND] .
\end{equation}

We recall that in statistical mechanics the partition function is
given by
\begin{eqnarray}
Z(\beta ) = \int dE \rho(E) \exp[-\beta E],
\end{eqnarray}
where $\rho(E)$ is the density of states that have energy $E$, and
$\beta$ is the inverse temperature.  Thus the subextensive entropy in
our learning problem is analogous to a system in which energy
corresponds to the Kullback--Leibler divergence relative to the target
model, and temperature is inverse to the number of examples.  As we
increase the length $N$ of the time series we have observed, we
``cool'' the system and hence probe models which approach the target;
the dynamics of this approach is determined by the density of low
energy states, that is the behavior of $\rho (D;{\bar{\bgm\alpha}})$
as $D\rightarrow 0$.\footnote{It may be worth emphasizing that this
analogy to statistical mechanics emerges from the analysis of
the relevant probability distributions, rather than being
imposed on the problem through some assumptions about the
nature of the learning algorithm.}

The structure of the partition function is determined by a competition
between the (Boltzmann) exponential term, which favors models with
small $D$, and the density term, which favors values of $D$ that can
be achieved by the largest possible number of models. Because there
(typically) are many parameters, there are very few models with $D
\rightarrow 0$.  This picture of competition between the Boltzmann
factor and a density of states has been emphasized in previous work on
supervised learning (Haussler et al. 1996).

The behavior of the density of states, $\rho (D;{\bar{\bgm\alpha}})$,
at small $D$ is related to the more intuitive notion of
dimensionality. In a parameterized family of distributions, the
Kullback--Leibler divergence between two distributions with nearby
parameters is approximately a quadratic form,
\begin{eqnarray}
D_{\rm KL} ( \bar{\bgm\alpha} || {\bgm\alpha} ) \approx {1\over 2}
\sum_{\mu\nu} (\bar{\alpha}_\mu  -\alpha_\mu) {\cal F}_{\mu\nu}
(\bar{\alpha}_\nu  -\alpha_\nu) + \cdots ,
\end{eqnarray}
where $N{\cal F}$ is the Fisher information matrix.  Intuitively, if we
have a reasonable parameterization of the distributions, then similar
distributions will be nearby in parameter space, and more importantly
points that are far apart in parameter space will never correspond to
similar distributions; Clarke and Barron (1990) refer to this
condition as the parameterization forming a ``sound'' family of
distributions.  If this condition is obeyed, then we can approximate
the low $D$ limit of the density $\rho (D;{\bar{\bgm\alpha}})$:
\begin{eqnarray}
\rho (D;{\bar{\bgm\alpha}}) &=& \int d^K\alpha {\cal P} ({\bgm\alpha})
\delta [D - D_{\rm KL} ( \bar{\bgm\alpha} || {\bgm\alpha} )]
\nonumber
\\
&\approx& \int d^K\alpha {\cal P} ({\bgm\alpha}) \delta \left[D - {1\over
2} \sum_{\mu\nu} (\bar{\alpha}_\mu  -\alpha_\mu) {\cal F}_{\mu\nu}
(\bar{\alpha}_\nu  -\alpha_\nu)\right]
\nonumber
\\ &=& \int d^K\alpha
{\cal P} (\bar{\bgm\alpha} + {\cal U}\cdot{\bgm\xi}) \delta\left[ D -
{1\over 2} \sum_\mu \Lambda_\mu\xi_\mu^2\right] ,
\label{rhoxi}
\end{eqnarray}
where $\cal U$ is a matrix that diagonalizes $\cal F$,
\begin{eqnarray}
({\cal U}^T \cdot {\cal F} \cdot {\cal U})_{\mu\nu} = \Lambda_\mu
\delta_{\mu\nu} .
\end{eqnarray}
The delta function restricts the components of $\bgm\xi$ in
Eq.~(\ref{rhoxi}) to be of order $\sqrt{D}$ or less, and so if
$P({\bgm\alpha})$ is smooth we can make a perturbation expansion.
After some algebra the leading term becomes
\begin{eqnarray}
\rho (D\rightarrow 0;{\bar{\bgm\alpha}}) \approx
{\cal P} ({\bar{\bgm\alpha}}) \frac{2 \pi^{K/2}}{\Gamma(K/2)} 
\left(\det \cal F\right)^{-1/2} D^{(K-2)/2} . 
\label{rho-finiteKclass}
\end{eqnarray}
Here, as before, $K$ is the dimensionality of the parameter vector.
Computing the partition function from Eq.~(\ref{ZD}), we find
\begin{equation}
Z ({\bar{\bgm\alpha}} ;N\rightarrow\infty) \approx
f({\bar{\bgm\alpha}}) \cdot {{\Gamma(K/2)}\over{N^{K/2}}} ,
\label{Zfinite}
\end{equation}
where $f({\bar{\bgm\alpha}})$ is some function of the target parameter
values. Finally, this allows us to evaluate the subextensive entropy,
from Eqs. (\ref{S1annealed}, \ref{Z}):
\begin{eqnarray}
S^{({\rm a})}_1(N) &=& - \int d^K{\bar\alpha} {\cal P}
(\bar{\bgm\alpha}) \log_2
Z(\bar{\bgm\alpha} ; N )
\label{s1thruZ}
\\
&\rightarrow& {K\over2}
\log_2 N + \cdots\;\;\; {\rm (bits)},
\label{s1distr}
\end{eqnarray}
where $\cdots$ are finite as $N \rightarrow \infty$. Thus, general
$K$--parameter model classes have the same subextensive entropy as for
the simplest example considered in the previous section.  To the
leading order, this result is independent even of the prior
distribution ${\cal P} ({\bgm\alpha})$ on the parameter space, so that
the predictive information seems to count the number of parameters
under some very general conditions [cf.\ Fig.~(\ref{entr_subext}) for
a very different numerical example of the logarithmic behavior].

Although Eq.~(\ref{s1distr}) is true under a wide range of conditions,
this cannot be the whole story.  Much of modern learning theory is
concerned with the fact that counting parameters is not quite enough
to characterize the complexity of a model class; the naive dimension
of the parameter space $K$ should be viewed in conjunction with the
pseudodimension (also known as the shattering dimension or
Vapnik--Chervonenkis dimension $d_{\rm VC}$), which measures capacity
of the model class, and with the phase space dimension $d$, which
accounts for volumes in the space of models (Vapnik 1998, Opper 1994).
Both of these dimensions can differ from the number of parameters. One
possibility is that $d_{\rm VC}$ is infinite when the number of
parameters is finite, a problem discussed below.  Another possibility
is that the determinant of ${\cal F}$ is zero, and hence $d_{\rm VC}$
and $d$ both are smaller than the number of parameters because we have
adopted a redundant description.  This sort of degeneracy could occur
over a finite fraction but not all of the parameter space, and this is
one way to generate an effective fractional dimensionality.  One can
imagine multifractal models such that the effective dimensionality
varies continuously over the parameter space, but it is not obvious
where this would be relevant.  Finally, models with $d<d_{\rm
  VC}<\infty$ also are possible [see, for example, Opper (1994)], and
this list probably is not exhaustive.

The calculation above, Eq.~(\ref{rho-finiteKclass}), lets us actually
{\em define} the phase space dimension through the exponent in the
small $D_{\rm KL}$ behavior of the model density,
\begin{eqnarray}
\rho(D\rightarrow 0; \bgm{\bar\alpha}) \propto D^{(d-2)/2} ,
\end{eqnarray}
and then $d$ appears in place of $K$ as the coefficient of the log
divergence in $S_1(N)$ (Clarke and Barron 1990, Opper 1994). However,
this simple conclusion can fail in two ways. First, it can happen that
the density $\rho(D;\bgm{\bar\alpha})$ accumulates a macroscopic
weight at some nonzero value of $D$, so that the small $D$ behavior is
irrelevant for the large $N$ asymptotics.  Second, the fluctuations
neglected here may be uncontrollably large, so that the asymptotics
are never reached. Since controllability of fluctuations is a function
of $d_{\rm VC}$ (see Vapnik 1998, Haussler et al.~1994, and later in
this paper), we may summarize this in the following way.  Provided
that the small $D$ behavior of the density function is the relevant
one, the coefficient of the logarithmic divergence of $I_{\rm pred}$
measures the phase space or the scaling dimension $d$ and nothing
else. This asymptote is valid, however, only for $N \gg d_{\rm VC}$.
It is still an open question whether the two pathologies that can
violate this asymptotic behavior are related.

\subsection{Learning a parameterized process}
\label{learn_proc_sec}

Consider a process where samples are not independent, and our task is
to learn their joint distribution $Q( {\vec x}_1 , \cdots, {\vec
  x}_{\rm N}| {\bgm\alpha})$. Again, ${\bgm\alpha}$ is an unknown
parameter vector which is chosen randomly at the beginning of the
series. If ${\bgm\alpha}$ is a $K$ dimensional vector, then one still
tries to learn just $K$ numbers and there are still $N$ examples, even
if there are correlations. Therefore, although such problems are much
more general than those considered above, it is reasonable to expect
that the predictive information still is measured by $(K/2)\log_2 N$
provided that some conditions are met.

One might suppose that conditions for simple results on the predictive
information are very strong, for example that the distribution $Q$ is
a finite order Markov model.  In fact all we really need are the
following two conditions:
\begin{eqnarray}
S\left[\{ {\vec x}_{\rm i}\}|{\bgm\alpha}\right] &\equiv& - \int
d^N{\vec x}\, Q(\{{\vec x}_{\rm i}\} | {\bgm\alpha})\, \log_2
Q(\{{\vec x}_{\rm i}\}| {\bgm\alpha})\nonumber\\
 &\to& N 
{\cal S}_0 + {\cal S}^*_0; \;\;\;\;\;\;\; {\cal S}^*_0 = O(1)\;
\label{S0markov},
\\ 
D_{\rm KL} \left[ Q(\{{\vec x}_{\rm i}\} |
\bar{\bgm\alpha})||Q(\{{\vec x}_{\rm i}\} | {\bgm\alpha})\right]
&\to & N {\cal D}_{\rm KL}\left( \bar{\bgm\alpha}||{\bgm\alpha}\right) +
o(N)\;.
\label{Dmarkov}
\end{eqnarray}
Here the quantities ${\cal S}_0$, ${\cal S}^*_0$, and ${\cal D}_{\rm
  KL}\left( \bar{\bgm\alpha}||{\bgm\alpha}\right)$ are defined by
taking limits $N\to\infty$ in both equations.  The first of the
constraints limits deviations from extensivity to be of order unity,
so that if ${\bgm\alpha}$ is known there are no long range
correlations in the data---all of the long range predictability is
associated with learning the parameters.\footnote{Suppose that we
  observe a Gaussian stochastic process and we try to learn the power
  spectrum.  If the class of possible spectra includes ratios of
  polynomials in the frequency (rational spectra) then this condition
  is met.  On the other hand, if the class of possible spectra
  includes $1/f$ noise, then the condition may not be met.  For more
  on long range correlations, see below.}  The second constraint,
Eq.~(\ref{Dmarkov}), is a less restrictive one, and it ensures that
the ``energy'' of our statistical system is an extensive quantity.

With these conditions it is straightforward to show that the results
of the previous subsection carry over virtually unchanged. With the
same cautious statements about fluctuations and the distinction
between $K$, $d$, and $d_{\rm VC}$, one arrives at the result:
\begin{eqnarray}
S(N) &=& {\cal S}_0 \cdot N + S^{({\rm a})}_1(N)\;,
\\
S^{({\rm a})}_1(N)&=& \frac{K}{2} \log_2 N + \cdots \;\;\;
({\rm bits}) ,
\label{s1mark}
\end{eqnarray}
where $\cdots$ stands for terms of order one as $N \to \infty$. Note
again that for the results Eq.~(\ref{s1mark}) to be valid, the process
considered is not required to be a finite order Markov process.
Memory of all previous outcomes may be kept, provided that the
accumulated memory does not contribute a divergent term to the
subextensive entropy.

It is interesting to ask what happens if the condition in
Eq.~(\ref{S0markov}) is violated, so that there are long range
correlations even in the conditional distribution $Q( {\vec x}_1 ,
\cdots, {\vec x}_{\rm N}| {\bgm\alpha})$.  Suppose, for example, that
${\cal S}^*_0=(K^*/2) \log_2 N$. Then the subextensive entropy becomes
\begin{equation}
S^{({\rm a})}_1(N)= \frac{K+K^*}{2} \log_2 N + \cdots \;\;\;
({\rm bits}) .
\label{s1markcorr}
\end{equation}
We see the that the subextensive entropy makes no distinction between
predictability that comes from unknown parameters and predictability
that comes from intrinsic correlations in the data; in this sense, two
models with the same $K+K^*$ are equivalent.  This, actually, {\em
  must} be so.  As an example, consider a chain of Ising spins with
long range interactions in one dimension.  This system can order
(magnetize) and exhibit long range correlations, and so the predictive
information will diverge at the transition to ordering.  In one view,
there is no global parameter analogous to $\bgm \alpha$, just the long
range interactions. On the other hand, there are regimes in which we
can approximate the effect of these interactions by saying that all
the spins experience a mean field which is constant across the whole
length of the system, and then formally we can think of the predictive
information as being carried by the mean field itself.  In fact there
are situations in which this is not just an approximation, but an
exact statement.  Thus we can trade a description in terms of long
range interactions ($K^* \neq 0$, but $K=0$) for one in which there
are unknown parameters describing the system but given these
parameters there are no long range correlations ($K\neq 0, \, K^*
=0$).  The two descriptions are equivalent, and this is captured by
the subextensive entropy.\footnote{There are a number of interesting
  questions about how the coefficients in the diverging predictive
  information relate to the usual critical exponents, and we hope to
  return to this problem in a later paper.}

\subsection{Taming the fluctuations}

The preceding calculations of the subextensive entropy $S_{\rm 1}$ are
worthless unless we prove that the fluctuations $\psi$ are
controllable. In this subsection we are going to discuss when and if
this, indeed, happens. We limit the discussion to the case of finding
a probability density (Section \ref{learn_distr_sec}); the case of
learning a process (Section \ref{learn_proc_sec}) is very similar.

Clarke and Barron (1990) solved essentially the same problem. They did
not make a separation into the annealed and the fluctuation term, and
the quantity they were interested in was a bit different from ours,
but, interpreting loosely, they proved that, modulo some reasonable
technical assumptions on differentiability of functions in question,
the fluctuation term always approaches zero. However, they did not
investigate the speed of this approach, and we believe that, by doing
so, they missed some important qualitative distinctions between
different problems that can arise due to a difference between $d$ and
$d_{\rm VC}$. In order to illuminate these distinctions, we here go
through the trouble of analyzing fluctuations all over again.

Returning to Eqs.~(\ref{pfn}, \ref{fn}) and the definition of entropy,
we can write the entropy $S(N)$ exactly as
\begin{eqnarray}
S(N)&=& - \int d^K\bar{\alpha} {\cal P} (\bar{\bgm\alpha})\int \prod_{j=1}^N 
\left[ d {\vec x}_{\rm j}\,
Q({\vec x}_{\rm j} | \bar{\bgm\alpha})\right]\nonumber\\
&&\times \log_2 \left[ \prod_{i=1}^N Q({\vec x}_{\rm i}|\bar{\bgm
\alpha}) 
\int  d^K\alpha
{\cal P} ({\bgm\alpha})\; {\rm e}^{-ND_{\rm KL}
(\bar{\bgms\alpha}||{\bgms\alpha}) + N \psi({\bgms
\alpha},\bar{\bgms\alpha};\{{\vec x}_{\rm i}\})}\right] .\nonumber\\
&&
\end{eqnarray}
This expression can be decomposed into the terms identified above,
plus a new contribution to the subextensive entropy that comes from
the fluctuations alone, $S_1^{\rm (f)}(N)$:
\begin{eqnarray}
S(N) &=& {\cal S}_0 \cdot N + S_1^{\rm (a)}(N) + S_1^{\rm (f)}(N),
\\
S^{({\rm f})}_1 &=&  - \int d^K {\bar\alpha}
{\cal P} (\bar{\bgm\alpha})
\,\prod_{j=1}^N \left[ 
d{\vec x}_{\rm j} \,Q({\vec x}_{\rm j}|\bar{\bgm \alpha})\right]
\nonumber
\\
&&\times\log_2 \left[ \int \frac{d^K\alpha
{\cal P} ({\bgm\alpha})}{Z(\bar{\bgm\alpha};N)} {\rm e}^{-ND_{\rm KL}
(\bar{\bgms\alpha}||{\bgms\alpha}) + N \psi({\bgms
\alpha},\bar{\bgms\alpha};\{{\vec x}_{\rm i}\})}\right] ,
\label{Sf}
\end{eqnarray}
where $\psi$ is defined as in Eq.~(\ref{fn}), and $Z$ as in
Eq.~(\ref{Z}).

Some loose but useful bounds can be established.  First, the
predictive information is a positive (semidefinite) quantity, and so
the fluctuation term may not be smaller (more negative) than the value
of $- S_{\rm 1}^{\rm (a)}$ as calculated in Eqs. (\ref{s1distr},
\ref{s1mark}).  Second, since fluctuations make it more difficult to
generalize from samples, the predictive information should {\em
  always} be reduced by fluctuations, so that $S^{({\rm f})}$ is
negative.  This last statement corresponds to the fact that for the
statistical mechanics of disordered systems, the annealed free energy
always is less than the average quenched free energy, and may be
proven rigorously by applying Jensen's inequality to the (concave)
logarithm function in Eq.~(\ref{Sf}); essentially the same argument
was given by Haussler and Opper (1997).  A related Jensen's inequality
argument allows us to show that the total $S_1(N)$ is bounded,
\begin{eqnarray}
S_1(N) &\leq& N \int d^K \alpha \int d^K \bar\alpha  
{\cal P} ({\bgm \alpha}) {\cal P} ({\bgm {\bar\alpha}}) 
D_{\rm KL} ( {\bgm{\bar\alpha}}||{\bgm\alpha}) 
\nonumber
\\
&\equiv& 
\langle N D_{\rm KL} ( {\bgm{\bar\alpha}}||{\bgm\alpha})  
\rangle_{{\bgms{\bar\alpha}},{\bgms\alpha}},
\label{boundbymeanKL}
\end{eqnarray}
so that if we have a class of models (and a prior ${\cal P} ({\bgm
  \alpha})$) such that the average Kullback--Leibler divergence among
pairs of models is finite, then the subextensive entropy necessarily
is properly defined. In its turn, finiteness of the average KL
divergence or of similar quantities is a usual constraint in the
learning problems of this type [see, for example, Haussler and Opper
(1997)]. Note also that $\langle D_{\rm KL}
({\bgm{\bar\alpha}}||{\bgm\alpha})
\rangle_{{\bgms{\bar\alpha}},{\bgms\alpha}}$ includes ${\cal S}_0$ as
one of its terms, so that usually ${\cal S}_0$ and $S_1$ are well-- or
ill--defined together.

Tighter bounds require nontrivial assumptions about the class of
distributions considered. The fluctuation term would be zero if $\psi$
were zero, and $\psi$ is the difference between an expectation value
(KL divergence) and the corresponding empirical mean. There is a broad
literature that deals with this type of difference (see, for example,
Vapnik 1998).

We start with the case when the pseudodimension ($d_{\rm VC}$) of the
set of probability densities $\{Q({\vec x}|{\bgm\alpha})\}$ is finite.
Then for any reasonable function $F({\vec x}; \beta)$, deviations of
the empirical mean from the expectation value can be bounded by
probabilistic bounds of the form
\begin{eqnarray}
&&P\left\{ \sup_{\beta} \left| \frac{\frac{1}{N} \sum_{\rm j}
F({\vec x}_{\rm j}; \beta) - \int d {\vec x}\, Q({\vec
x}|\bar{\bgm \alpha}) \,F({\vec x}; \beta)}{L[F]} \right| >
\epsilon \right\}
\nonumber
\\
&&\,\,\,\,\,\,\;\;\;\;\;\,\,\,\,\,\,
< M(\epsilon, N,d_{\rm VC}) {\rm e}^{-c
N\epsilon^2}\, ,
\label{GC}
\end{eqnarray}
where $c$ and $L[F]$ depend on the details of the particular bound
used.  Typically, $c$ is a constant of order one, and $L[F]$ either is
some moment of $F$ or the range of its variation. In our case, $F$ is
the log--ratio of two densities, so that $L[F]$ may be assumed bounded
for almost all $\beta$ without loss of generality in view of
Eq.~(\ref{boundbymeanKL}).  In addition, $M(\epsilon, N, d_{\rm VC})$
is finite at zero, grows at most subexponentially in its first two
arguments, and depends exponentially on $d_{\rm VC}$.  Bounds of this
form may have different names in different contexts:
Glivenko--Cantelli, Vapnik--Chervonenkis, Hoeffding, Chernoff, ...;
for review see Vapnik (1998) and the references therein.

To start the proof of finiteness of $S^{({\rm f})}_1$ in this case, we
first show that only the region ${\bgm\alpha}\approx\bar{\bgm\alpha}$
is important when calculating the inner integral in Eq.~(\ref{Sf}).
This statement is equivalent to saying that at large values of
${\bgm\alpha} - \bar{\bgm\alpha}$ the KL divergence almost always
dominates the fluctuation term, that is, the contribution of sequences
of $\{{\vec x}_{\rm i}\}$ with atypically large fluctuations is
negligible (atypicality is defined as $\psi\ge \delta$, where $\delta$
is some small constant independent of $N$). Since the fluctuations
decrease as $1/\sqrt{N}$ [see Eq.~(\ref{GC})], and $D_{\rm KL}$ is of
order one, this is plausible. To show this, we bound the logarithm in
Eq.~(\ref{Sf}) by $N$ times the supremum value of $\psi$.  Then we
realize that the averaging over $\bar{\bgm\alpha}$ and $\{{\vec
  x}_{\rm i}\}$ is equivalent to integration over all possible values
of the fluctuations. The worst case density of the fluctuations may be
estimated by differentiating Eq.~(\ref{GC}) with respect to $\epsilon$
(this brings down an extra factor of $N\epsilon$). Thus the worst case
contribution of these atypical sequences is
\begin{eqnarray}
S^{({\rm f}),{\rm atypical}}_1 \sim \int_{\delta}^{\infty}
d\epsilon\, N^2 \epsilon^2 M(\epsilon) {\rm e}^{-c N \epsilon^2}
\sim {\rm e}^{-cN\delta^2} \ll 1\; {\rm for \;large\; } N .
\end{eqnarray}

This bound lets us focus our attention on the region
${\bgm\alpha}\approx\bar{\bgm\alpha}$. We expand the exponent of the
integrand of Eq.~(\ref{Sf}) around this point and perform a simple
Gaussian integration. In principle, large fluctuations might lead to
an instability (positive or zero curvature) at the saddle point, but
this is atypical and therefore is accounted for already. Curvatures at
the saddle points of both numerator and denominator are of the same
order, and throwing away unimportant additive and multiplicative
constants of order unity, we obtain the following result for the
contribution of typical sequences:
\begin{eqnarray}
S^{({\rm f}),{\rm typical}}_1 &\sim& \int d^K {\bar\alpha}
{\cal P} (\bar{\bgm\alpha}) \,d^N{\vec x} \prod_{\rm j} Q({\vec x}_{\rm
j}|\bar{\bgm \alpha})\; N\; ({\bf B} \,{\cal A}^{-1}\, {\bf
B})\;;\label{chi2}
\\
B_{\mu}&=&\frac{1}{N}\sum_{\rm i} \frac{\partial \log Q({\vec x}_{\rm
i}|\bar{\bgm\alpha})}{\partial
\bar{\alpha}_{\mu}}\;,\;\;\; \langle{\bf B}\rangle_{\vec x}=0\;;
\nonumber
\\
({\cal A})_{\mu \nu}&=&\frac{1}{N}\sum_{\rm i}
\frac{\partial^2 \log Q({\vec x}_{\rm
i}|\bar{\bgm\alpha})}{\partial \bar{\alpha}_{\mu} \partial
\bar{\alpha}_{\nu}} \;,\;\;\; \langle{\cal A}\rangle_{\vec x}={\cal
F}\;.
\nonumber
\end{eqnarray}
Here $\langle \cdots \rangle_{\vec x}$ means an averaging with respect
to all ${\vec x}_{\rm i}$'s keeping $\bar{\bgm\alpha}$ constant. One
immediately recognizes that ${\bf B}$ and ${\cal A}$ are,
respectively, first and second derivatives of the empirical KL
divergence that was in the exponent of the inner integral in
Eq.~(\ref{Sf}).

We are dealing now with typical cases. Therefore, large deviations of
${\cal A}$ from ${\cal F}$ are not allowed, and we may bound
Eq.~(\ref{chi2}) by replacing ${\cal A}^{-1}$ with ${\cal F}^{-1}
(1+\delta)$, where $\delta$ again is independent of $N$.  Now we have
to average a bunch of products like
\begin{eqnarray}
\frac{\partial \log Q({\vec x}_{\rm i}|\bar{\bgm \alpha})}
{\partial \bar{\alpha}_{\mu}} ({\cal F}^{-1})_{\mu\nu} \frac{\partial
  \log Q({\vec x}_{\rm j}|\bar{\bgm\alpha})} {\partial
  \bar{\alpha}_{\nu}}
\end{eqnarray}
over all ${\vec x}_{\rm i}$'s. Only the terms with ${\rm i}={\rm j}$
survive the averaging. There are $K^2N$ such terms, each contributing
of order $N^{-1}$. This means that the total contribution of the
typical fluctuations is bounded by a number of order one and does not
grow with $N$.

This concludes the proof of controllability of fluctuations for
$d_{\rm VC}<\infty$. One may notice that we never used the specific
form of $M(\epsilon,N,d_{\rm VC})$, which is the only thing dependent
on the precise value of the dimension.  Actually, a more thorough look
at the proof shows that we do not even need the strict uniform
convergence enforced by the Glivenko--Cantelli bound. With some
modifications the proof should still hold if there exist some {\em a
  priori} improbable values of ${\bgm\alpha}$ and $\bar{\bgm\alpha}$
that lead to violation of the bound. That is, if the prior ${\cal P}
({\bgm\alpha})$ has sufficiently narrow support, then we may still
expect fluctuations to be unimportant even for VC--infinite problems.

A proof of this can be found in the realm of the Structural Risk
Minimization (SRM) theory (Vapnik 1998). SRM theory assumes that an
infinite {\em structure} ${\mathcal C}$ of nested subsets $C_1 \subset
C_2 \subset C_3 \subset \cdots$ can be imposed onto the set $C$ of all
admissible solutions of a learning problem, such that $C= \bigcup
C_{\rm n}$. The idea is that, having a finite number of observations
$N$, one is confined to the choices within some particular structure
element $C_{\rm n},\, {\rm n}={\rm n}(N),$ when looking for an
approximation to the true solution; this prevents overfitting and poor
generalization. Then, as the number of samples increases and one is
able to distinguish within more and more complicated subsets, $\rm n$
grows.  If $d_{\rm VC}$ for learning in any $C_{\rm n},\, {\rm n} <
\infty,$ is finite, then one can show convergence of the estimate to
the true value as well as the absolute smallness of fluctuations
(Vapnik 1998). It is remarkable that this result holds even if the
capacity of the whole set $C$ is not described by a finite $d_{\rm
  VC}$.

In the context of SRM, the role of the prior $P({\bgm\alpha})$ is to
induce a structure on the set of all admissible densities, and the
fight between the number of samples $N$ and the narrowness of the
prior is precisely what determines how the capacity of the current
element of the structure $C_{\rm n},\, {\rm n}={\rm n}(N),$ grows with
$N$. A rigorous proof of smallness of the fluctuations can be
constructed based on well known results, as detailed elsewhere
(Nemenman 2000).  Here we focus on the question of how narrow the
prior should be so that every structure element is of finite
VC--dimension, and one can guarantee eventual convergence of
fluctuations to zero.

Consider two examples. A variable $x$ is distributed according to the
following probability density functions:
\begin{eqnarray}
Q(x|\alpha) &=& \frac{1}{\sqrt{2\pi}} \exp \left[ -\frac{1}{2}
\left(x-\alpha\right)^2\right]\;, \;\; x\in (-\infty;+\infty)\;;
\\
Q(x|\alpha) &=& \frac{\exp \left( - \sin \alpha x
  \right)}{\int_0^{2\pi} dx\,\exp \left( - \sin \alpha x
  \right)}\;,\;\;
x\in [0; 2\pi)\;.
\label{expsin}
\end{eqnarray}
Learning the parameter in the first case is a $d_{\rm VC}=1$ problem,
while in the second case $d_{\rm VC}=\infty$.  In the first example,
as we have shown above, one may construct a uniform bound on
fluctuations irrespective of the prior $P({\bgm\alpha})$. The second
one does not allow this. Suppose that the prior is uniform in a box
$0<\alpha < \alpha_{\rm max}$, and zero elsewhere, with $\alpha_{\rm
  max}$ rather large.  Then for not too many sample points $N$, the
data would be better fitted not by some value in the vicinity of the
actual parameter, but by some much larger value, for which almost all
data points are at the crests of $-\sin \alpha x$. Adding a new data
point would not help, until that best, but wrong, parameter estimate
is less than $\alpha_{\rm max}$.\footnote{Interestingly, since for the
  model Eq.~(\ref{expsin}) KL divergence is bounded from below and
  from {\em above}, for $\alpha_{\rm max} \to \infty$ the weight in
  $\rho(D;{\bar{\bgm\alpha}})$ at small $D_{\rm KL}$ vanishes, and a
  finite weight accumulates at some nonzero value of $D$. Thus, even
  putting the fluctuations aside, the asymptotic behavior based on the
  phase space dimension is invalidated, as mentioned above.} So the
fluctuations are large, and the predictive information is small in
this case.  Eventually, however, data points would overwhelm the box
size, and the best estimate of $\alpha$ would swiftly approach the
actual value. At this point the argument of Clarke and Barron would
become applicable, and the leading behavior of the subextensive
entropy would converge to its asymptotic value of $(1/2) \log N$. On
the other hand, there is no uniform bound on the value of $N$ for
which this convergence will occur---it is guaranteed only for $N\gg
d_{\rm VC}$, which is never true if $d_{\rm VC}=\infty$. For some
sufficiently wide priors this asymptotically correct behavior would be
never reached in practice. Further, if we imagine a thermodynamic
limit where the box size and the number of samples both become large,
then by analogy with problems in supervised learning (Seung et al.\ 
1992, Haussler et al.\ 1996) we expect that there can be sudden
changes in performance as a function of the number of examples.  The
arguments of Clarke and Barron cannot encompass these phase
transitions or ``aha!'' phenomena. A further bridge between VC
dimension and the information theoretic approach to learning may be
found in Haussler et al.\ (1994), where the authors bounded predictive
information--like quantities with loose but useful bounds explicitly
proportional to $d_{\rm VC}$.

While much of learning theory has focused on problems with finite VC
dimension, it might be that the conventional scenario in which the
number of examples eventually overwhelms the number of parameters or
dimensions is too weak to deal with many real world problems.
Certainly in the present context there is not only a quantitative, but
also a qualitative difference between reaching the asymptotic regime
in just a few measurements, or in many millions of them. Finitely
parameterizable models with finite or infinite $d_{\rm VC}$ fall in
{\em essentially different universality classes} with respect to the
predictive information.

\subsection{Beyond finite parameterization: \\general considerations}
\label{nonparam_gen_sec}

The previous sections have considered learning from time series where
the underlying class of possible models is described with a finite
number of parameters.  If the number of parameters is not finite then
in principle it is impossible to learn anything unless there is some
appropriate regularization of the problem.  If we let the number of
parameters stay finite but become large, then there is {\em more} to
be learned and correspondingly the predictive information grows in
proportion to this number, as in Eq.~(\ref{s1distr}). On the other
hand, if the number of parameters becomes infinite without
regularization, then the predictive information should go to zero
since nothing can be learned.  We should be able to see this happen in
a regularized problem as the regularization weakens: eventually the
regularization would be insufficient and the predictive information
would vanish.  The only way this can happen is if the subextensive
term in the entropy grows more and more rapidly with $N$ as we weaken
the regularization, until finally it becomes extensive at the point
where learning becomes impossible. More precisely, if this scenario
for the breakdown of learning is to work, there must be situations in
which the predictive information grows with $N$ more rapidly than the
logarithmic behavior found in the case of finite parameterization.

Subextensive terms in the entropy are controlled by the density of
models as a  function of their Kullback--Leibler divergence to the target
model.  If the models have finite VC and phase space dimensions then
this density vanishes for small divergences as $\rho \sim
D^{(d-2)/2}$.  Phenomenologically, if we let the number of parameters
increase, the density vanishes more and more rapidly. We can imagine
that beyond the class of finitely parameterizable problems there is a
class of regularized infinite dimensional problems in which the
density $\rho (D \rightarrow 0)$ vanishes more rapidly than any power
of $D$.  As an example, we could have
\begin{eqnarray}
\rho (D \rightarrow 0) \approx A \exp\left[
-{B\over{D^\mu}}\right] ,\,\,\,\,\,\,\, \mu>0 ;
\end{eqnarray}
that is, an essential singularity at $D=0$. For simplicity we assume
that the constants $A$ and $B$ can depend on the target model, but
that the nature of the essential singularity ($\mu$) is the same
everywhere.  Before providing an explicit example, let us explore the
consequences of this behavior.

From Equation (\ref{ZD}) above, we can write the partition function as
\begin{eqnarray}
Z (\bar{\bgm\alpha} ; N ) &=& \int dD \rho (D;{\bar{\bgm\alpha}})
\exp[-ND]
\nonumber
\\
&\approx& A (\bar{\bgm\alpha}) \int dD\exp \left[
-{{B(\bar{\bgm\alpha}) }\over{D^\mu}} - ND \right]
\nonumber
\\
&\approx&
{\tilde A}(\bar{\bgm\alpha}) \exp\left[ -{1\over 2}
{{\mu+2}\over{\mu+1}}\ln N - C(\bar{\bgm\alpha}) N^{\mu/(\mu+1)}
\right] ,
\label{Zinfgeneral}
\end{eqnarray}
where in the last step we use a saddle point or steepest descent
approximation which is accurate at large $N$, and the coefficients are
\begin{eqnarray}
{\tilde A}(\bar{\bgm\alpha}) &=& A(\bar{\bgm\alpha}) \left({{2\pi
\mu^{1/(\mu+1)}} \over{\mu+1}} \right)^{1/2}\cdot
[B(\bar{\bgm\alpha})]^{1/(2\mu+2)}
\\
C(\bar{\bgm\alpha}) &=& [B(\bar{\bgm\alpha})]^{1/(\mu+1)} \left(
{1\over{\mu^{\mu/(\mu+1)}}} + \mu^{1/(\mu+1)} \right).
\end{eqnarray}
Finally we can use Eqs.~(\ref{s1thruZ}, \ref{Zinfgeneral}) to compute
the subextensive term in the entropy, keeping only the dominant term
at large $N$,
\begin{equation}
S_1^{({\rm a})} (N) \rightarrow {1\over{\ln 2}} 
\langle C (\bar{\bgm\alpha})
\rangle_{\bar{\bgms\alpha}} N^{\mu/(\mu+1)} \;\;\; ({\rm bits}),
\label{s1power}
\end{equation}
where $\langle \cdots \rangle_{\bar{\bgms\alpha}}$ denotes an average
over all the target models.

This behavior of the first subextensive term is qualitatively
different from everything we have observed so far. A power law
divergence is much stronger than a logarithmic one. Therefore, a lot
more predictive information is accumulated in an ``infinite
parameter'' (or nonparametric) system; the system is much richer and
more complex, both intuitively and quantitatively.

Subextensive entropy also grows as a power law in a finitely
parameterizable system with a growing number of parameters.  For
example, suppose that we approximate the distribution of a random
variable by a histogram with $K$ bins, and we let $K$ grow with the
quantity of available samples as $K \sim N^\nu$.  A straightforward
generalization of Eq.~(\ref{s1distr}) seems to imply then that $S_1(N)
\sim N^{\nu} \log N$ (Hall and Hannan 1988, Barron and Cover 1991).
While not necessarily wrong, analogies of this type are dangerous. If
the number of parameters grows constantly, then the scenario where the
data overwhelms all the unknowns is far from certain. Observations may
provide much less information about features that were
introduced into the model at some large $N$ than about those that have
existed since the very first measurements. Indeed, in a $K$--parameter
system, the $N^{\rm th}$ sample point contributes $\sim K/2N$ bits to
the subextensive entropy [cf.\ Eq.~(\ref{s1distr})]. If $K$ changes as
mentioned, the $N^{\rm th}$ example then carries $\sim N^{\nu-1}$
bits. Summing this up over all samples, we find $S^{({\rm a})}_1 \sim
N^{\nu}$, and if we let $\nu = \mu/(\mu +1)$ we obtain
Eq.~(\ref{s1power}); note that the growth of the number of parameters
is slower than $N$ ($\nu = \mu/(\mu +1) < 1$), which makes sense.
Rissanen et al.\ (1992) made a similar observation. According to them,
for models with increasing number of parameters, predictive codes,
which are optimal at each particular $N$ [cf.\ 
Eq.~(\ref{condentropy})], provide a strictly shorter coding length
than nonpredictive codes optimized for all data simultaneously.  This
has to be contrasted with the finite parameter model classes, for
which these codes are asymptotically equal.

Power law growth of the predictive information illustrates the point
made earlier about the transition from learning more to finally
learning nothing as the class of investigated models becomes more
complex. As $\mu$ increases, the problem becomes richer and more
complex, and this is expressed in the stronger divergence of the first
subextensive term of the entropy; for fixed large $N$, the predictive
information increases with $\mu$.  However, if $\mu\rightarrow\infty$
the problem is too complex for learning---in our model example the
number of bins grows in proportion to the number of samples, which
means that we are trying to find too much detail in the underlying
distribution.  As a result, the subextensive term becomes extensive
and stops contributing to predictive information. Thus, at least to
the leading order, predictability is lost, as promised.

\subsection{Beyond finite parameterization: example}
\label{nonparam_ex_sec}

While literature on problems in the logarithmic class is reasonably
rich, the research on establishing the power law behavior seems to be
in its early stages.  Some authors have found specific learning
problems for which quantities similar to, but sometimes very
nontrivially related to $S_1$, are bounded by power--law functions
(Haussler and Opper 1997, 1998, Cesa--Bianchi and Lugosi 2000).
Others have chosen to study finite dimensional models, for which the
optimal number of parameters [usually determined by the MDL criterion
of Rissanen (1989)] grows as a power law (Hall and Hannan 1988,
Rissanen et al.~1992). In addition to the questions raised earlier
about the danger of copying finite dimensional intuition to the
infinite dimensional setup, these are not examples of truly
nonparametric Bayesian learning. Instead these authors make use of a
priori constraints to restrict learning to codes of particular
structure (histogram codes), while a non--Bayesian inference is
conducted within the class. Without Bayesian averaging and with
restrictions on the coding strategy it may happen that a realization
of the code length is substantially different from the predictive
information.  Similar conceptual problems plague even true
nonparametric examples, as considered, for example, by Barron and
Cover (1991). In summary, we don't know of a complete calculation in
which a family of power--law scalings of the predictive information is
derived from a Bayesian formulation.

The discussion in the previous section suggests that we should look
for power--law behavior of the predictive information in learning
problems where rather than learning ever more precise values for a
fixed set of parameters, we learn a progressively more detailed
description---effectively increasing the number of parameters---as we
collect more data. One example of such a problem is learning the
distribution $Q(x)$ for a continuous variable $x$, but rather than
writing a parametric form of $Q(x)$ we assume only that this function
itself is chosen from some distribution that enforces a degree of
smoothness. There are some natural connections of this problem to the
methods of quantum field theory (Bialek, Callan, and Strong 1996)
which we can exploit to give a complete calculation of the predictive
information, at least for a class of smoothness constraints.

We write $Q(x)= (1/l_0) \exp [-\phi(x)]$ so that positivity of the
distribution is automatic, and then smoothness may be expressed by
saying that the `energy' (or action) associated with a function
$\phi(x)$ is related to an integral over its derivatives, like the
strain energy in a stretched string.  The simplest possibility
following this line of ideas is that the distribution of functions is
given by
\begin{equation}
\label{prior_nonparam}
{\cal P}[\phi(x)]= \frac{1}{{\cal Z}} \exp \left[-\frac{l}{2}\int dx
\left(\frac{\partial \phi}{\partial x}\right)^2\right] \delta
\left[\frac{1}{l_0}\int dx\, {\rm e}^{- \phi(x)} -1 \right]\; ,
\end{equation}
where ${\cal Z}$ is the normalization constant for ${\cal P}[\phi]$,
the delta function insures that each distribution $Q(x)$ is
normalized, and $l$ sets a scale for smoothness. If distributions are
chosen from this distribution, then the optimal Bayesian estimate of
$Q(x)$ from a set of samples $x_1 , x_2, \cdots , x_N$ converges to
the correct answer, and the distribution at finite $N$ is nonsingular,
so that the regularization provided by this prior is strong enough to
prevent the development of singular peaks at the location of observed
data points (Bialek, Callan, and Strong 1996). Further developments of
the theory, including alternative choices of $P[\phi(x)]$, have been
given by Periwal (1997, 1998), Holy (1997) and Aida (1998); for a
detailed numerical investigation of this problem see Nemenman and
Bialek (2001).  Our goal here is to be illustrative rather than
exhaustive.\footnote{We caution the reader that our discussion in this
  section is less self--contained than in other sections.  Since the
  crucial steps exactly parallel those in the earlier work, here we
  just give references.}

From the discussion above we know that the predictive information is
related to the density of Kullback--Leibler divergences, and that the
power--law behavior we are looking for comes from an essential
singularity in this density function.  Thus we calculate
$\rho(D,\bar{\phi})$ in the model defined by
Eq.~(\ref{prior_nonparam}).

With $Q(x)= (1/l_0) \exp [-\phi(x)]$, we can write the KL divergence
as
\begin{equation}
D_{\rm KL} [\bar\phi (x) ||\phi (x) ]
= {1\over{l_0}}\int dx \exp[-\bar\phi(x)] [\phi(x) - \bar\phi(x)]\, .
\end{equation}
We want to compute the density,
\begin{eqnarray}
\rho (D; \bar\phi) &=& \int [d\phi(x)] {\cal P}[\phi(x)]
\delta\left( D - D_{\rm KL} [\bar\phi (x) ||\phi (x) ]\right)\\
&=&
M \int [d\phi(x)] {\cal P} [\phi(x)]
\delta\left( MD - MD_{\rm KL} [\bar\phi (x) ||\phi (x) ]\right),
\end{eqnarray}
where we introduce a factor $M$ which we will allow to become large so
that we can focus our attention on the interesting limit $D\rightarrow
0$.  To compute this integral over all functions $\phi(x)$, we
introduce a Fourier representation for the delta function, and then
rearrange the terms:
\begin{eqnarray}
\rho (D; \bar\phi) &=&
M \int {{dz}\over{2\pi}} \exp(izMD) \int [d\phi(x)] {\cal P} [\phi(x)]
\exp(-izMD_{\rm KL})\\
&=&
M \int {{dz}\over{2\pi}} \exp\left(izMD
+ {{izM}\over{l_0}} \int dx \bar\phi(x)\exp[-\bar\phi(x)]\right)
\nonumber\\
&&\times \int [d\phi(x)] {\cal P} [\phi(x)]\exp\left(
-{{izM}\over{l_0}}\int dx \phi(x)\exp[-\bar\phi(x)]
\right) .
\end{eqnarray}
The inner integral over the functions $\phi(x)$ is exactly the
integral which was evaluated in the original discussion of this
problem (Bialek, Callan and Strong 1996); in the limit that $zM$ is
large we can use a saddle point approximation, and standard field
theoretic methods allow us to compute the fluctuations around the
saddle point.  The result is that
$$
\int [d\phi(x)] {\cal P}[\phi(x)]\exp\left( -{{izM}\over{l_0}}\int
  dx \phi(x)\exp[-\bar\phi(x)] \right)
$$
\begin{eqnarray}
&=&  
\exp\left(
-{{izM}\over{l_0}}\int dx \bar\phi(x)\exp[-\bar\phi(x)]
- S_{\rm eff}[\bar\phi(x);zM]\right),\nonumber\\
&&\\
S_{\rm eff}[\bar\phi;zM]&=&
\frac{l}{2}\int dx
\left(\frac{\partial \bar\phi}{\partial x}\right)^2
+\frac{1}{2} \left({{izM}\over{l l_0}}\right)^{1/2} 
\int dx \exp[-\bar\phi(x)/2] .
\nonumber\\
&&\end{eqnarray} 
Now we can do the integral over $z$, again by a saddle point method.
The two saddle point approximations are both valid in the limit that
$D\rightarrow 0$ and $MD^{3/2} \rightarrow\infty$; we are interested
precisely in the first limit, and we are free to set $M$ as we wish,
so this gives us a good approximation for $\rho(D\rightarrow 0 ;
\bar\phi)$.  This results in 
\begin{eqnarray}
\rho(D\to 0 ; \bar\phi) &=& A[\bar\phi(x)] D^{-3/2}
\exp\left( - {{B [\bar\phi(x)]}\over D}\right) ,\\
A[\bar\phi(x)]
&=&
{1\over{\sqrt{16\pi l l_0}}}
\exp\left[-\frac{l}{2}\int dx
\left(\partial_x \bar\phi \right)^2\right]
\int {{dx}}
\exp[-\bar\phi(x)/2 ]
\\
B [\bar\phi(x)]
&=&
{1\over{16 l l_0}}
\left(\int {{dx}}
\exp[-\bar\phi(x)/2]\right)^2 \, .
\end{eqnarray}
Except for the factor of $D^{-3/2}$, this is exactly the sort of
essential singularity that we considered in the previous section, with
$\mu =1$.  The $D^{-3/2}$ prefactor does not change the leading large
$N$ behavior of the predictive information, and we find that
\begin{equation}
S^{({\rm a})}_{\rm 1}(N) \sim
{1\over{2\ln 2\sqrt{l l_0}}}
\Bigg\langle
\int dx \exp[-\bar\phi(x)/2]\Bigg\rangle_{\bar\phi}
N^{1/2} ,
\label{almostdone}
\end{equation}
where $\langle \cdots \rangle_{\bar\phi}$ denotes an average over the
target distributions $\bar\phi(x)$ weighted once again by ${\cal
  P}[\bar\phi(x)]$.  Notice that if $x$ is unbounded then the average
in Eq.~(\ref{almostdone}) is infrared divergent; if instead we let the
variable $x$ range from $0$ to $L$ then this average should be
dominated by the uniform distribution.  Replacing the average by its
value at this point, we obtain the approximate result
\begin{equation}\label{S1infbox}
S^{({\rm a})}_{\rm 1}(N)\sim \frac{1}{2\ln 2} \,\sqrt{N}\,
\left(\frac{L}{l}\right)^{1/2} \,{\rm bits.}
\end{equation}

To understand the result in Eq.~(\ref{S1infbox}), we recall that this
field theoretic approach is more or less equivalent to an adaptive
binning procedure in which we divide the range of $x$ into bins of
local size $\sqrt{l/NQ(x)}$ (Bialek, Callan, and Strong 1996).  From
this point of view, Eq.~(\ref{S1infbox}) makes perfect sense: the
predictive information is directly proportional to the number of bins
that can be put in the range of $x$. This also is in direct accord
with a comment from the previous subsection that power law behavior of
predictive information arises from the number of parameters in the
problem depending on the number of samples.

This counting of parameters allows a schematic argument about the
smallness of fluctuations in this particular nonparametric problem. If
we take the hint that at every step $\sim \sqrt{N}$ bins are being
investigated, then we can imagine that the field theoretic prior in
Eq.~(\ref{prior_nonparam}) has imposed a structure ${\mathcal C}$ on
the set of all possible densities, so that the set $C_{\rm n}$ is
formed of all continuous piecewise linear functions that have not more
than ${\rm n}$ kinks. Learning such functions for finite ${\rm n}$ is
a $d_{\rm VC}={\rm n}$ problem. Now, as $N$ grows, the elements with
higher capacities ${\rm n}\sim \sqrt{N}$ are admitted. The
fluctuations in such a problem are known to be controllable (Vapnik
1998), as discussed in more detail elsewhere (Nemenman 2000).

One thing which remains troubling is that the predictive information
depends on the arbitrary parameter $l$ describing the scale of
smoothness in the distribution.  In the original work it was proposed
that one should integrate over possible values of $l$ (Bialek, Callan
and Strong 1996).  Numerical simulations demonstrate that this
parameter can be learned from the data itself (Nemenman and Bialek
2000), but perhaps even more interesting is a formulation of the
problem by Periwal (1997, 1998) which recovers complete coordinate
invariance by effectively allowing $l$ to vary with $x$.  In this case
the whole theory has no length scale, and there also is no need to
confine the variable $x$ to a box (here of size $L$).  We expect that
this coordinate invariant approach will lead to a universal
coefficient multiplying $\sqrt{N}$ in the analog of
Eq.~(\ref{S1infbox}), but this remains to be shown.

In summary, the field theoretic approach to learning a smooth
distribution in one dimension provides us with a concrete, calculable
example of a learning problem with power--law growth of the predictive
information.  The scenario is exactly as suggested in the previous
section, where the density of KL divergences develops an essential
singularity.  Heuristic considerations (Bialek, Callan, and Strong
1996; Aida 1999) suggest that different smoothness penalties and
generalizations to higher dimensional problems will have sensible
effects on the predictive information---all have power--law growth,
smoother functions have smaller powers (less to learn), and higher
dimensional problems have larger powers (more to learn)---but real
calculations for these cases remain challenging.

\section{$I_{\rm pred}$ as a measure of complexity}

The problem of quantifying complexity is very old [see Grassberger
(1991) for a short review].  Solomonoff (1964), Kolmogorov (1965), and
Chaitin (1975) investigated a mathematically rigorous notion of
complexity that measures (roughly) the minimum length of a computer
program that simulates the observed time series [see also Li and
Vit{\'a}nyi (1993)].  Unfortunately there is no algorithm that can
calculate the Kolmogorov complexity of all data sets. In addition,
algorithmic or Kolmogorov complexity is closely related to the Shannon
entropy, which means that it measures something closer to our
intuitive concept of randomness than to the intuitive concept of
complexity [as discussed, for example, by Bennett (1990)].  These
problems have fueled continued research along two different paths,
representing two major motivations for defining complexity.  First,
one would like to make precise an impression that some systems---such
as life on earth or a turbulent fluid flow---evolve toward a state of
higher complexity, and one would like to be able to classify those
states. Second, in choosing among different models that describe an
experiment, one wants to quantify a preference for simpler
explanations or, equivalently, provide a penalty for complex models
that can be weighed against the more conventional goodness of fit
criteria. We bring our readers up to date with some developments in
both of these directions, and then discuss the role of predictive
information as a measure of complexity. This also gives us an
opportunity to discuss more carefully the relation of our results to
previous work.

\subsection{Complexity of statistical models}

The construction of complexity penalties for model selection is a
statistics problem. As far as we know, however, the first discussions
of complexity in this context belong to philosophical literature. Even
leaving aside the early contributions of William of Occam on the need
for simplicity, Hume on the problem of induction, and Popper on
falsifiability, Kemeney (1953) suggested explicitly that it would be
possible to create a model selection criterion that balances goodness
of fit versus complexity. Wallace and Burton (1968) hinted that this
balance may result in the model with ``the briefest recording of all
attribute information.'' Even though he probably had a somewhat
different motivation, Akaike (1974a, 1974b) made the first
quantitative step along these lines. His {\em ad hoc} complexity term
was independent of the number of data points and was proportional to
the number of free independent parameters in the model.

These ideas were rediscovered and developed systematically by Rissanen
in a series of papers starting from 1978. He has emphasized strongly
(Rissanen 1984, 1986, 1987, 1989) that fitting a model to data
represents an encoding of those data, or predicting future data, and
that in searching for an efficient code we need to measure not only
the number of bits required to describe the deviations of the data
from the model's predictions (goodness of fit), but also the number of
bits required to specify the parameters of the model (complexity).
This specification has to be done to a precision supported by the
data.\footnote{Within this framework Akaike's suggestion can be seen
  as coding the model to (suboptimal) fixed precision.}  Rissanen
(1984) and Clarke and Barron (1990) in full generality were able to
prove that the optimal encoding of a model requires a code with length
asymptotically proportional to the number of independent parameters
and logarithmically dependent on the number of data points we have
observed.  The minimal amount of space one needs to encode a data
string (minimum description length or MDL) within a certain assumed
model class was termed by Rissanen {\em stochastic complexity,} and in
recent work he refers to the piece of the stochastic complexity
required for coding the parameters as the {\em model complexity}
(Rissanen 1996). This approach was further strengthened by a recent
result (Vit{\'a}nyi and Li 2000) that an estimation of parameters
using the MDL principle is equivalent to Bayesian parameter
estimations with a ``universal'' prior (Li and Vit{\'a}nyi 1993).

There should be a close connection between Rissanen's ideas of
encoding the data stream and the subextensive entropy.  We are
accustomed to the idea that the average length of a code word for
symbols drawn from a distribution $P$ is given by the entropy of that
distribution; thus it is tempting to say that an encoding of a stream
$x_1 , x_2, \cdots , x_N$ will require an amount of space equal to the
entropy of the joint distribution $P(x_1 , x_2, \cdots , x_N )$.  The
situation here is a bit more subtle, because the usual proofs of
equivalence between code length and entropy rely on notions of
typicality and asymptotics as we try to encode sequences of many
symbols; here we already have $N$ symbols and so it doesn't really
make sense to talk about a stream of streams. One can argue, however,
that atypical sequences are not truly random within a considered
distribution since their coding by the methods optimized for the
distribution is not optimal. So atypical sequences are better
considered as typical ones coming from a different distribution [a
point also made by Grassberger (1986)]. This allows us to identify
properties of an observed (long) string with the properties of the
distribution it comes from, as was done by Vit{\'a}nyi and Li (2000).
If we accept this identification of entropy with code length, then
Rissanen's stochastic complexity should be the entropy of the
distribution $P(x_1 , x_2, \cdots , x_N)$.

As emphasized by Balasubramanian (1997), the entropy of the joint
distribution of $N$ points can be decomposed into pieces that
represent noise or errors in the model's local predictions---an
extensive entropy---and the space required to encode the model itself,
which must be the subextensive entropy.  Since in the usual
formulation all long--term predictions are associated with the
continued validity of the model parameters, the dominant component of
the subextensive entropy must be this parameter coding, or model
complexity in Rissanen's terminology.  Thus the subextensive entropy
should be the model complexity, and in simple cases where we can
describe the data by a $K$--parameter model both quantities are equal
to $(K/2)\log_2 N$ bits to the leading order.

The fact that the subextensive entropy or predictive information
agrees with Rissanen's model complexity suggests that $I_{\rm pred}$
provides a reasonable measure of complexity in learning problems.
This agreement might lead the reader to wonder if all we have done is
to rewrite the results of Rissanen et al.~in a different notation. To
calm these fears we recall again that our approach distinguishes
infinite VC problems from finite ones and treats nonparametric cases
as well. Indeed, the predictive information is defined without
reference to the idea that we are learning a model, and thus we can
make a link to physical aspects of the problem, as discussed below.

The MDL principle was introduced as a procedure for statistical
inference from a data stream to a model.  In contrast, we take the
predictive information to be a {\em characterization} of the data
stream itself.  To the extent that we can think of the data stream as
arising from a model with unknown parameters, as in the examples of
Section 4, all notions of inference are purely Bayesian and there is
no additional ``penalty for complexity.''  In this sense our
discussion is much closer in spirit to Balasubramanian (1997) than to
Rissanen (1978).  On the other hand, we can always think about the
ensemble of data streams that can be generated by a {\em class} of
models, provided that we have a proper Bayesian prior on the members
of the class.  Then the predictive information measures the complexity
of the class, and this characterization can be used to understand why
inference within some (simpler) model classes will be more efficient;
for practical examples along these lines see Nemenman (2000) and
Nemenman and Bialek (2001).

\subsection{Complexity of dynamical systems}

While there are a few attempts to quantify complexity in terms of
deterministic predictions (Wolfram 1984), the majority of efforts to
measure the complexity of physical systems starts with a probabilistic
approach.  In addition, there is a strong prejudice that the
complexity of physical systems should be measured by quantities that
are not only statistical, but are also at least related to more
conventional thermodynamic quantities (temperature, entropy, $\dots$),
since this is the only way one will be able to calculate complexity
within the framework of statistical mechanics. Most proposals define
complexity as an entropy--like quantity, but an entropy of some
unusual ensemble.  For example, Lloyd and Pagels (1988) identified
complexity as {\em thermodynamic depth}, the entropy of the state
sequences that lead to the current state. The idea clearly is in the
same spirit as the measurement of predictive information, but this
depth measure does not completely discard the extensive component of
the entropy (Crutchfield and Shalizi 1999) and thus fails to resolve
the essential difficulty in constructing complexity measures for
physical systems: distinguishing genuine complexity from randomness
(entropy), the complexity should be zero both for purely regular and
for purely random systems.

New definitions of complexity that try to satisfy these criteria
(Lopez--Ruiz et al.~1995, Gell--Mann and Lloyd 1996, Shiner et
al.~1999, Sole and Luque 1999, Adami and Cerf 2000) and criticisms of
these proposals (Crutchfield et al.~1999, Feldman and Crutchfield
1998, Sole and Luque 1999) continue to emerge even now. Aside from the
obvious problems of not actually eliminating the extensive component
for all or a part of the parameter space or not expressing complexity
as an average over a physical ensemble, the critiques often are based
on a clever argument first mentioned explicitly by Feldman and
Crutchfield (1998). In an attempt to create a universal measure, the
constructions can be made {\em over--universal}: many proposed
complexity measures depend only on the entropy density ${\cal S}_0$
and thus are functions only of disorder---not a desired feature. In
addition, many of these and other definitions are flawed because they
fail to distinguish among the richness of classes beyond some very
simple ones.

In a series of papers, Crutchfield and coworkers identified {\em
  statistical complexity} with the entropy of {\em causal states,}
which in turn are defined as all those microstates (or histories) that
have the same conditional distribution of futures (Crutchfield and
Young 1989, Shalizi and Crutchfield 1999).  The causal states provide
an optimal description of a system's dynamics in the sense that these
states make as good a prediction as the histories themselves.
Statistical complexity is very similar to predictive information, but
Shalizi and Crutchfield (1999) define a quantity which is even closer
to the spirit of our discussion: their {\em excess entropy} is exactly
the mutual information between the semi--infinite past and future.
Unfortunately, by focusing on cases in which the past and future are
infinite but the excess entropy is finite, their discussion is limited
to systems for which (in our language) $I_{\rm
  pred}(T\rightarrow\infty) = {\rm constant}$.

In our view, Grassberger (1986, 1991) has made the clearest and the
most appealing definitions. He emphasized that the slow approach of
the entropy to its extensive limit is a sign of complexity, and has
proposed several functions to analyze this slow approach. His {\em
  effective measure complexity} is the subextensive entropy term of an
infinite data sample.  Unlike Crutchfield et al., he allows this
measure to grow to infinity. As an example, for low dimensional
dynamical systems, the effective measure complexity is finite whether
the system exhibits periodic or chaotic behavior, but at the
bifurcation point that marks the onset of chaos, it diverges
logarithmically. More interestingly, Grassberger also notes that
simulations of specific cellular automaton models that are capable of
universal computation indicate that these systems can exhibit an even
stronger, power--law, divergence.

Grassberger (1986, 1991) also introduces the {\em true measure
  complexity}, or the {\em forecasting complexity}, which is the
minimal information one needs to extract from the past in order to
provide optimal prediction. Interestingly, another complexity measure,
the {\em logical depth} (Bennett 1985), which measures the time needed
to decode the optimal compression of the observed data, is bounded
from below by this quantity because decoding requires
reading all of this information. In addition, true measure
complexity is exactly the statistical complexity of Crutchfield et
al., and the two approaches are actually much closer than they seem.
The relation between the true and the effective measure complexities,
or between the statistical complexity and the excess entropy, closely
parallels the idea of extracting or compressing relevant information
(Tishby et al.~1999, Bialek and Tishby, in preparation), as discussed
below.

\subsection{A unique measure of complexity?}

We recall that entropy provides a measure of information that is
unique in satisfying certain plausible constraints (Shannon 1948). It
would be attractive if we could prove a similar uniqueness theorem for
the predictive information, or any part of it, as a measure of the
complexity or richness of a time dependent signal $x(0 < t <T)$ drawn
from a distribution $P[x(t)]$. Before proceeding along these lines we
have to be more specific about what we mean by ``complexity.''  In
most cases, including the learning problems discussed above, it is
clear that we want to measure complexity of the dynamics underlying
the signal, or equivalently the complexity of a model that might be
used to describe the signal.\footnote{The problem of finding this
  model or of reconstructing the underlying dynamics may also be
  complex in the computational sense, so that there may not exist an
  efficient algorithm.  More interestingly, the computational effort
  required may grow with the duration $T$ of our observations.  We
  leave these algorithmic issues aside for the present discussion.}
There remains a question, however, whether we want to attach measures
of complexity to a particular signal $x(t)$, or whether we are
interested in measures (like the entropy itself) that are defined by
an average over the ensemble $P[x(t)]$.

One problem in assigning complexity to single realizations is that
there can be atypical data streams.  Either we must treat atypicality
explicitly, arguing that atypical data streams from one source should
be viewed as typical streams from another source, as discussed by
Vit{\'a}nyi and Li (2000), or we have to look at average quantities.
Grassberger (1986) in particular has argued that our visual intuition
about the complexity of spatial patterns is an ensemble concept, even
if the ensemble is only implicit; see also Tong in the discussion
session of Rissanen (1987).  In Rissanen's formulation of MDL, one
tries to compute the description length of a single string with
respect to some class of possible models for that string, but if these
models are probabilistic we can always think about these models as
generating an ensemble of possible strings.  The fact that we admit
probabilistic models is crucial: even at a colloquial level, if we
allow for probabilistic models then there is a simple description for
a sequence of truly random bits, but if we insist on a deterministic
model then it may be very complicated to generate precisely the
observed string of bits.\footnote{This is the statement that the
  Kolmogorov complexity of a random sequence is large: the programs or
  algorithms considered in the Kolmogorov formulation are
  deterministic, and the program must generate precisely the observed
  string.}  Furthermore, in the context of probabilistic models it
hardly makes sense to ask for a dynamics that generates a particular
data stream; we must ask for dynamics that generate the data with
reasonable probability, which is more or less equivalent to asking
that the given string be a typical member of the ensemble generated by
the model.  All of these paths lead us to thinking not about single
strings but about ensembles in the tradition of statistical mechanics,
and so we shall search for measures of complexity that are averages
over the distribution $P[x(t)]$.

Once we focus on average quantities, we can start by adopting
Shannon's postulates as constraints on a measure of complexity: if
there are $N$ equally likely signals, then the measure should be
monotonic in $N$; if the signal is decomposable into statistically
independent parts then the measure should be additive with respect to
this decomposition; and if the signal can be described as a leaf on a
tree of statistically independent decisions then the measure should be
a weighted sum of the measures at each branching point. We believe
that these constraints are as plausible for complexity measures as for
information measures, and it is well known from Shannon's original
work that this set of constraints leaves the entropy as the only
possibility.  Since we are discussing a time dependent signal, this
entropy depends on the duration of our sample, $S(T)$.  We know of
course that this cannot be the end of the discussion, because we need
to distinguish between randomness (entropy) and complexity.  The path
to this distinction is to introduce other constraints on our measure.

First we notice that if the signal $x$ is continuous, then the entropy
is not invariant under transformations of $x$.  It seems reasonable to
ask that complexity be a function of the process we are observing and
not of the coordinate system in which we choose to record our
observations. The examples above show us, however, that it is not the
whole function $S(T)$ which depends on the coordinate system for
$x$;\footnote{Here we consider instantaneous transformations of $x$,
  not filtering or other transformations that mix points at different
  times.} it is only the extensive component of the entropy that has
this noninvariance.  This can be seen more generally by noting that
subextensive terms in the entropy contribute to the mutual information
among different segments of the data stream (including the predictive
information defined here), while the extensive entropy cannot; mutual
information is coordinate invariant, so all of the noninvariance must
reside in the extensive term.  Thus, any measure complexity that is
coordinate invariant must discard the extensive component of the
entropy.

The fact that extensive entropy cannot contribute to complexity is
discussed widely in the physics literature (Bennett 1990), as our
short review above shows. To statisticians and computer scientists,
who are used to Kolmogorov's ideas, this is less obvious.  However,
Rissanen (1986, 1987) also talks about ``noise'' and ``useful
information'' in a data sequence, which is similar to splitting
entropy into its extensive and the subextensive parts. His ``model
complexity,'' aside from not being an average as required above, is
essentially equal to the subextensive entropy. Similarly, Whittle [in
the discussion of Rissanen (1987)] talks about separating the
predictive part of the data from the rest.

If we continue along these lines, we can think about the asymptotic
expansion of the entropy at large $T$.  The extensive term is the
first term in this series, and we have seen that it must be discarded.
What about the other terms?  In the context of learning a
parameterized model, most of the terms in this series depend in detail
on our prior distribution in parameter space, which might seem odd for
a measure of complexity.  More generally, if we consider
transformations of the data stream $x(t)$ that mix points within a
temporal window of size $\tau$, then for $T >> \tau$ the entropy
$S(T)$ may have subextensive terms which are constant, and these are
not invariant under this class of transformations.  On the other hand,
if there are divergent subextensive terms, these {\em are} invariant
under such temporally local transformations.\footnote{Throughout this
  discussion we assume that the signal $x$ at one point in time is
  finite dimensional.  There are subtleties if we allow $x$ to
  represent the configuration of a spatially infinite system.}  So if
we insist that measures of complexity be invariant not only under
instantaneous coordinate transformations, but also under temporally
local transformations, then we can discard both the extensive and the
finite subextensive terms in the entropy, leaving only the divergent
subextensive terms as a possible measure of complexity.

An interesting example of these arguments is provided by the
statistical mechanics of polymers.  It is conventional to make models
of polymers as random walks on a lattice, with various interactions or
self avoidance constraints among different elements of the polymer
chain.  If we count the number $\cal N$ of walks with $N$ steps, we
find that ${\cal N}(N) \sim A N^\gamma z^N$ (de Gennes 1979). Now the
entropy is the logarithm of the number of states, and so there is an
extensive entropy ${\cal S}_0 = \log_2 z$, a constant subextensive
entropy $\log_2 A$, and a divergent subextensive term $S_1(N)
\rightarrow \gamma \log_2 N$.  Of these three terms, only the
divergent subextensive term (related to the critical exponent
$\gamma$) is universal, that is independent of the detailed structure
of the lattice.  Thus, as in our general argument, it is only the
divergent subextensive terms in the entropy that are invariant to
changes in our description of the local, small scale dynamics.

We can recast the invariance arguments in a slightly different form
using the relative entropy.  We recall that entropy is defined cleanly
only for discrete processes, and that in the continuum there are
ambiguities.  We would like to write the continuum generalization of
the entropy of a process $x(t)$ distributed according to $P[x(t)]$ as
\begin{eqnarray}
S_{\rm cont} = -\int Dx(t) \,P[x(t)]\log_2 P[x(t)] ,
\end{eqnarray}
but this is not well defined because we are taking the logarithm of a
dimensionful quantity.  Shannon gave the solution to this problem: we
use as a measure of information the relative entropy or KL divergence
between the distribution $P[x(t)]$ and some reference distribution
$Q[x(t)]$,
\begin{eqnarray}
S_{\rm rel} = -\int Dx(t) \,P[x(t)]\log_2
\left({{P[x(t)]}\over{Q[x(t)]}} \right) ,
\end{eqnarray}
which is invariant under changes of our coordinate system on the space
of signals.  The cost of this invariance is that we have introduced an
arbitrary distribution $Q[x(t)]$, and so really we have a family of
measures.  We can find a unique complexity measure within this family
by imposing invariance principles as above, but in this language we
must make our measure invariant to different choices of the reference
distribution $Q[x(t)]$.

The reference distribution $Q[x(t)]$ embodies our expectations for the
signal $x(t)$; in particular, $S_{\rm rel}$ measures the extra space
needed to encode signals drawn from the distribution $P[x(t)]$ if we
use coding strategies that are optimized for $Q[x(t)]$. If $x(t)$ is a
written text, two readers who expect different numbers of spelling
errors will have different $Q$s, but to the extent that spelling
errors can be corrected by reference to the immediate neighboring
letters we insist that any measure of complexity be invariant to these
differences in $Q$. On the other hand, readers who differ in their
expectations about the global subject of the text may well disagree
about the richness of a newspaper article. This suggests that
complexity is a component of the relative entropy that is invariant
under some class of local translations and misspellings.

Suppose that we leave aside global expectations, and construct our
reference distribution $Q[x(t)]$ by allowing only for short ranged
interactions---certain letters tend to follow one another, letters
form words, and so on, but we bound the range over which these rules
are applied. Models of this class cannot embody the full structure of
most interesting time series (including language), but in the present
context we are not asking for this.  On the contrary, we are looking
for a measure that is invariant to differences in this short ranged
structure.  In the terminology of field theory or statistical
mechanics, we are constructing our reference distribution $Q[x(t)]$
from local operators.  Because we are considering a one dimensional
signal (the one dimension being time), distributions constructed from
local operators cannot have any phase transitions as a function of
parameters; again it is important that the signal $x$ at one point in
time is finite dimensional.  The absence of critical points means that
the entropy of these distributions (or their contribution to the
relative entropy) consists of an extensive term (proportional to the
time window $T$) plus a constant subextensive term, plus terms that
vanish as $T$ becomes large.  Thus, if we choose different reference
distributions within the class constructible from local operators, we
can change the extensive component of the relative entropy, and we can
change constant subextensive terms, but the divergent subextensive
terms are invariant.

To summarize, the usual constraints on information measures in the
continu\-um produce a family of allowable complexity measures, the
relative entropy to an arbitrary reference distribution. If we insist
that all observers who choose reference distributions constructed from
local operators arrive at the same measure of complexity, or if we
follow the first line of arguments presented above, then this measure
must be the divergent subextensive component of the entropy or,
equivalently, the predictive information. We have seen that this
component is connected to learning in a straightforward way,
quantifying the amount that can be learned about dynamics that
generate the signal, and to measures of complexity that have arisen in
statistics and in dynamical systems theory.

\section{Discussion}
\label{pred_discus_sec}

We have presented predictive information as a {\em characterization}
of various data streams.  In the context of learning, predictive
information is related directly to generalization.  More generally,
the structure or order in a time series or a sequence is related
almost by definition to the fact that there is predictability along
the sequence.  The predictive information measures the amount of such
structure, but doesn't exhibit the structure in a concrete form.
Having collected a data stream of duration $T$, what are the features
of these data that carry the predictive information $I_{\rm pred}(T)$?
From Equation (\ref{chuck}) we know that most of what we have seen
over the time $T$ must be irrelevant to the problem of prediction, so
that the predictive information is a small fraction of the total
information; can we separate these predictive bits from the vast
amount of nonpredictive data?

The problem of separating predictive from nonpredictive information is
a special case of the problem discussed recently (Tishby et al.~1999,
Bialek and Tishby, in preparation): given some data $x$, how do we
compress our description of $x$ while preserving as much information
as possible about some other variable $y$?  Here we identify $x =
x_{\rm past}$ as the past data and $y= x_{\rm future}$ as the future.
When we compress $x_{\rm past}$ into some reduced description $\hat
x_{\rm past}$ we keep a certain amount of information about the past,
$I(\hat x_{\rm past} ; x_{\rm past})$, and we also preserve a certain
amount of information about the future, $I(\hat x_{\rm past} ; x_{\rm
  future})$.  There is no single correct compression $x_{\rm past}
\rightarrow \hat x_{\rm past}$; instead there is a one parameter
family of strategies which trace out an optimal curve in the plane
defined by these two mutual informations, $I(\hat x_{\rm past} ;
x_{\rm future})$ vs.  $I(\hat x_{\rm past} ; x_{\rm past})$.

The predictive information preserved by compression must be less than
the total, so that $I(\hat x_{\rm past} ; x_{\rm future}) \leq I_{\rm
  pred}(T)$.  Generically no compression can preserve all of the
predictive information so that the inequality will be strict, but
there are interesting special cases where equality can be achieved. If
prediction proceeds by learning a model with a finite number of
parameters, we might have a regression formula that specifies the best
estimate of the parameters given the past data; using the regression
formula compresses the data but preserves all of the predictive power.
In cases like this (more generally, if there exist sufficient
statistics for the prediction problem) we can ask for the minimal set
of $\hat x_{\rm past}$ such that $I(\hat x_{\rm past} ; x_{\rm
  future}) = I_{\rm pred}(T)$.  The entropy of this minimal $\hat
x_{\rm past}$ is the true measure complexity defined by Grassberger
(1986) or the statistical complexity defined by Crutchfield and Young
(1989) [in the framework of the causal states theory a very similar
comment was made recently by Shalizi and Crutchfield (2000)].

In the context of statistical mechanics, long range correlations are
characterized by computing the correlation functions of order
parameters, which are coarse--grained functions of the system's
microscopic variables.  When we know something about the nature of the
order parameter (e.~g., whether it is a vector or a scalar), then
general principles allow a fairly complete classification and
description of long range ordering and the nature of the critical
points at which this order can appear or change.  On the other hand,
defining the order parameter itself remains something of an art.  For
a ferromagnet, the order parameter is obtained by local averaging of
the microscopic spins, while for an antiferromagnet one must average
the staggered magnetization to capture the fact that the ordering
involves an alternation from site to site, and so on.  Since the order
parameter carries all the information that contributes to long range
correlations in space and time, it might be possible to define order
parameters more generally as those variables that provide the most
efficient compression of the predictive information, and this should
be especially interesting for complex or disordered systems where the
nature of the order is not obvious intuitively; a first try in this
direction was made by Bruder (1998). At critical points the predictive
information will diverge with the size of the system, and the
coefficients of these divergences should be related to the standard
scaling dimensions of the order parameters, but the details of this
connection need to be worked out.

If we compress or extract the predictive information from a time
series we are in effect discovering ``features'' that capture the
nature of the ordering in time.  Learning itself can be seen as an
example of this, where we discover the parameters of an underlying
model by trying to compress the information that one sample of $N$
points provides about the next, and in this way we address directly
the problem of generalization (Bialek and Tishby, in preparation).
The fact that (as mentioned above) nonpredictive information is
useless to the organism suggests that one crucial goal of neural
information processing is to separate predictive information from the
background.  Perhaps rather than providing an efficient representation
of the current state of the world---as suggested by Attneave (1954),
Barlow (1959, 1961), and others (Atick 1992)---the nervous system
provides an efficient representation of the predictive
information.\footnote{If, as seems likely, the stream of data reaching
  our senses has diverging predictive information then the space
  required to write down our description grows and grows as we observe
  the world for longer periods of time.  In particular, if we can
  observe for a very long time then the amount that we know about the
  future will exceed, by an arbitrarily large factor, the amount that
  we know about the present.  Thus representing the predictive
  information may require many more neurons than would be required to
  represent the current data.  If we imagine that the goal of primary
  sensory cortex is to represent the current state of the sensory
  world, then it is difficult to understand why these cortices have so
  many more neurons than they have sensory inputs.  In the extreme
  case, the region of primary visual cortex devoted to inputs from the
  fovea has nearly 30,000 neurons for each photoreceptor cell in the
  retina (Hawken and Parker 1991); although much remains to be learned
  about these cells, it is difficult to imagine that the activity of
  so many neurons constitutes an efficient representation of the
  current sensory inputs.  But if we live in a world where the
  predictive information in the movies reaching our retina diverges,
  it is perfectly possible that an efficient representation of the
  predictive information available to us at one instant requires
  thousands of times more space than an efficient representation of
  the image currently falling on our retina.}  It should be possible
to test this directly by studying the encoding of reasonably natural
signals and asking if the information which neural responses provide
about the future of the input is close to the limit set by the
statistics of the input itself, given that the neuron only captures a
certain number of bits about the past.  Thus we might ask if, under
natural stimulus conditions, a motion sensitive visual neuron captures
features of the motion trajectory that allow for optimal prediction or
extrapolation of that trajectory; by using information theoretic
measures we both test the ``efficient representation'' hypothesis
directly and avoid arbitrary assumptions about the metric for errors
in prediction.  For more complex signals such as communication sounds,
even identifying the features that capture the predictive information
is an interesting problem.

It is natural to ask if these ideas about predictive information could
be used to analyze experiments on learning in animals or humans.  We
have emphasized the problem of learning probability distributions or
probabilistic models rather than learning deterministic functions,
associations or rules.  It is known that the nervous system adapts to
the statistics of its inputs, and this adaptation is evident in the
responses of single neurons (Smirnakis et al.~1996, Brenner et
al.~2000); these experiments provide a simple example of the system
learning a parameterized distribution.  When making saccadic eye
movements, human subjects alter their distribution of reaction times
in relation to the relative probabilities of different targets, as if
they had learned an estimate of the relevant likelihood ratios
(Carpenter and Williams 1995).  Humans also can learn to discriminate
almost optimally between random sequences (fair coin tosses) and
sequences that are correlated or anticorrelated according to a Markov
process; this learning can be accomplished from examples alone, with
no other feedback (Lopes and Oden 1987).  Acquisition of language may
require learning the joint distribution of successive phonemes,
syllables, or words, and there is direct evidence for learning of
conditional probabilities from artificial sound sequences, both by
infants and by adults (Saffran et al.~1996; 1999).  These examples,
which are not exhaustive, indicate that the nervous system can learn
an appropriate probabilistic model,\footnote{As emphasized above, many
  other learning problems, including learning a function from noisy
  examples, can be seen as the learning of a probabilistic model.
  Thus we expect that this description applies to a much wider range
  of biological learning tasks.} and this offers the opportunity to
analyze the dynamics of this learning using information theoretic
methods: What is the entropy of $N$ successive reaction times
following a switch to a new set of relative probabilities in the
saccade experiment?  How much information does a single reaction time
provide about the relevant probabilities?  Following the arguments
above, such analysis could lead to a measurement of the universal
learning curve $\Lambda (N)$.

The learning curve $\Lambda (N)$ exhibited by a human observer is
limited by the predictive information in the time series of stimulus
trials itself.  Comparing $\Lambda (N)$ to this limit defines an
efficiency of learning in the spirit of the discussion by Barlow
(1983). While it is known that the nervous system can make efficient
use of available information in signal processing tasks [cf. Chapter 4
of Rieke et al.~(1997)], and that it can represent
this information efficiently in the spike trains of individual neurons
[cf. Chapter 3 of Rieke et al.~(1997), as well as
Berry, Warland and  Meister (1997), Strong et al. (1998),
and Reinagel and Reid (2000)], it is not known whether the brain is an
efficient learning machine in the analogous sense.  Given our
classification of learning tasks by their complexity, it would be
natural to ask if the efficiency of learning were a critical function
of task complexity: perhaps we can even identify a limit beyond which
efficient learning fails, indicating a limit to the complexity of the
internal model used by the brain during a class of learning tasks.  We
believe that our theoretical discussion here at least frames a clear
question about the complexity of internal models, even if for the
present we can only speculate about the outcome of such experiments.

An important result of our analysis is the characterization of time
series or learning problems beyond the class of finitely
parameterizable models, that is the class with power--law divergent
predictive information.  Qualitatively this class is more complex than
{\em any} parametric model, no matter how many parameters there may
be, because of the more rapid asymptotic growth of $I_{\rm pred}(N)$.
On the other hand, with a finite number of observations $N$, the
actual amount of predictive information in such a nonparametric
problem may be {\em smaller} than in a model with a large but finite
number of parameters.  Specifically, if we have two models, one with
$I_{\rm pred}(N)\sim AN^\nu$ and one with $K$ parameters so that
$I_{\rm pred}(N)\sim (K/2)\log_2 N$, the infinite parameter model has
less predictive information for all $N$ smaller than some critical
value
\begin{equation}
N_c \sim \left[ {K\over{2A\nu}}\log_2\left(
{K\over{2A}}\right)\right]^{1/\nu} .
\label{Nc}
\end{equation}
In the regime $N << N_c$, it is possible to achieve more efficient
prediction by trying to learn the (asymptotically) more complex model,
as illustrated concretely in simulations of the density estimation
problem (Nemenman and Bialek 2000).  Even if there are a finite number
of parameters---such as the finite number of synapses in a small
volume of the brain---this number may be so large that we always have
$N \ll N_c$, so that it may be more effective to think of the many
parameter model as approximating a continuous or nonparametric one.

It is tempting to suggest that the regime $N << N_c$ is the relevant
one for much of biology.  If we consider, for example, 10 mm$^2$ of
inferotemporal cortex devoted to object recognition (Logothetis and
Sheinberg 1996), the number of synapses is $K\sim 5\times 10^9$.  On
the other hand, object recognition depends on foveation, and we move
our eyes roughly three times per second throughout perhaps 10 years of
waking life during which we master the art of object recognition.
This limits us to at most $N\sim 10^9$ examples.  Remembering that we
must have $\nu < 1$, even with large values of $A$ Eq.~(\ref{Nc})
suggests that we operate with $N < N_c$.  One can make similar
arguments about very different brains, such as the mushroom bodies in
insects (Capaldi, Robinson and Fahrbach 1999). If this identification
of biological learning with the regime $N << N_c$ is correct, then the
success of learning in animals must depend on strategies that
implement sensible priors over the space of possible models.

There is one clear empirical hint that humans can make effective use
of models that are beyond finite parameterization (in the sense that
predictive information diverges as a power--law), and this comes from
studies of language.  Long ago, Shannon (1951) used the knowledge of
native speakers to place bounds on the entropy of written English, and
his strategy made explicit use of predictability.  Shannon showed
$N$--letter sequences to native speakers (readers), asked them to
guess the next letter, and recorded how many guesses were required
before they got the right answer.  Thus each letter in the text is
turned into a number, and the entropy of the distribution of these
numbers is an upper bound on the conditional entropy $\ell (N)$ [cf.
Eq.~(\ref{condentropy})].  Shannon himself thought that the
convergence as $N$ becomes large was rather quick, and quoted an
estimate of the extensive entropy per letter ${\cal S}_0$.  Many years
later, Hilberg (1990) reanalyzed Shannon's data and found that the
approach to extensivity in fact was very slow: certainly there is
evidence for a large component $S_1(N) \propto N^{1/2}$, and this may
even dominate the extensive component for accessible $N$.  Ebeling and
P{\"o}schel (1994; see also P\"oschel, Ebeling, and Ros\'e 1995)
studied the statistics of letter sequences in long texts (like {\em
  Moby Dick}) and found the same strong subextensive component.  It
would be attractive to repeat Shannon's experiments with a design that
emphasizes the detection of subextensive terms at large
$N$.\footnote{Associated with the slow approach to extensivity is a
  large mutual information between words or characters separated by
  long distances, and several groups have found that this mutual
  information declines as a power law. Cover and King (1978) criticize
  such observations by noting that this behavior is impossible in
  Markov chains of arbitrary order. While it is possible that existing
  mutual information data have not reached asymptotia, the criticism
  of Cover and King misses the possibility that language is {\em not}
  a Markov process.  Of course it cannot be Markovian if it has a
  power--law divergence in the predictive information.}

In summary, we believe that our analysis of predictive information
solves the problem of measuring the complexity of time series.  This
analysis unifies ideas from learning theory, coding theory, dynamical
systems, and statistical mechanics.  In particular we have focused
attention on a class of processes that are qualitatively more complex
than those treated in conventional learning theory, and there are
several reasons to think that this class includes many examples of
relevance to biology and cognition.

\subsection*{Acknowledgements}

We thank V.\ Balasubramanian, A.\ Bell, S.\ Bruder, C.\ Callan, A.\ 
Fairhall, G.\ Garcia de Polavieja Embid, R.\ Koberle, A.\ Libchaber,
A.\ Melikidze, A.\ Mikhailov, O.\ Motrunich, M.\ Opper, R.\ Rumiati,
R.\ de Ruyter van Steveninck, N.\ Slonim, T.\ Spencer, S.\ Still, S.\ 
Strong, and A.\ Treves for many helpful discussions.  We also thank
M.\ Nemenman and R.\ Rubin for a help with the numerical simulations,
and an anonymous referee for pointing out yet more opportunities to
connect our work with the earlier literature.  Our collaboration was
aided in part by a grant from the US--Israel Binational Science
Foundation to the Hebrew University of Jerusalem, and work at
Princeton was supported in part by funds from NEC.

\section{References}

\begin{description}
    \item Abarbanel, H.~D.~I., Brown, R., Sidorowich, J.~J., \&
  Tsimring L.~S.~(1993). The analysis of observed chaotic data in
  physical systems, {\em Revs.\ Mod.\ Phys.} {\bf 65}, 1331--1392.
    \item Adami, C., \& Cerf N.~J.\ (2000). Physical complexity of
  symbolic sequences, {\it Physica D} {\bf 137}, 62--69. See also
  adap-org/9605002.\footnote{Where available, we give references to
    the Los Alamos e--print archive. Papers may be retrieved from the
    web site {\em http://xxx.lanl.gov/abs/*/*}, where {\em */*} is the
    reference; thus Adami and Cerf (2000) is found at {\em
      http://xxx.lanl.gov/abs/adap-org/9605002}.  For preprints this
    is a primary reference; for published papers there may be
    differences between the published version and the e--print. }
    \item Aida, T.~(1999).  Field theoretical analysis of on--line
  learning of probability distributions, {\it Phys.\ Rev.\ Lett.} {\bf
    83}, 3554--3557.  See also cond-mat/9911474.
    \item Akaike, H.\ (1974a). Information theory and an extension of
  the maximum likelihood principle, in {\em Second international
    symposium of information theory}, B.~Pet\-rov and F.~Csaki, eds.
  Budapest: (Akademia Kiedo, Budapest).
    \item Akaike, H.~(1974b). A new look at the statistical model
  identification, {\em IEEE Trans.\ Automatic Control.} {\bf 19},
  716--723.
    \item Atick, J.~J.~(1992).  Could information theory provide an
  ecological theory of sensory processing?  In {\em Princeton Lectures
    on Biophysics}, W.~Bialek, ed., pp.~223--289 (World Scientific,
  Singapore).
    \item Attneave, F.~(1954).  Some informational aspects of visual
  perception, {\em Psych.\ Rev.} {\bf 61}, 183--193.
    \item Balasubramanian, V.~(1997).  Statistical inference, Occam's
  razor, and statistical mechanics on the space of probability
  distributions, {\em Neural Comp.} {\bf 9}, 349--368.  See also
  cond-mat/9601030.
    \item Barlow, H.~B.~(1959).  Sensory mechanisms, the reduction of
  redundancy and intelligence, in {\em Proceedings of the Symposium on
    the Mechanization of Thought Processes, vol.~2}, D.~V.~Blake and
  A.~M.~Uttley, eds., pp.~537--574 (H.~M.~Stationery Office, London).
    \item Barlow, H.~B.~(1961).  Possible principles underlying the
  transformation of sensory messages, in {\em Sensory Communication},
  W.~Rosenblith, ed., pp.~217--234 (MIT Press, Cambridge).
    \item Barlow, H.~B.~(1983).  Intelligence, guesswork, language,
  {\em Nature} {\bf 304}, 207--209.
    \item Barron, A., \& Cover, T.~(1991). Minimum complexity density
  estimation, {\em IEEE Trans.\ Inf.\ Thy.} {\bf 37}, 1034--1054.
    \item Bennett, C.~H.~(1985). Dissipation, information,
  computational complexity and the definition of organization, in {\em
    Emerging Syntheses in Science}, D.\ Pines, ed., pp.~215-233
  (Addison--Wesley, Reading, MA).
    \item Bennett, C.~H.~(1990).  How to define complexity in physics,
  and why, in {\it Complexity, Entropy and the Physics of
    Information}, W.~H.~Zurek, ed., pp.~137--148 (Addison--Wesley,
  Redwood City).
    \item Berry II, M.\ J., Warland, D.\ K., \& Meister, M.\ (1997).  The
  structure and precision of retinal spike trains, {\em Proc.\ Natl.\
    Acad.\  Sci.\ (USA)} {\bf 94,} 5411--5416.
    \item Bialek, W.~(1995).  Predictive information and the
  complexity of time series, NEC Research Institute technical note.
    \item Bialek, W., Callan, C.~G., \& Strong, S.~P.~(1996). Field
  theories for learning probability distributions, {\it
    Phys.~Rev.~Lett.}  {\bf 77}, 4693--4697.  See also\\
  cond-mat/9607180.
    \item Bialek, W., \& Tishby, N.~(1999). Predictive information,
  preprint.  Available at cond-mat/9902341.
    \item Bialek, W., \& Tishby, N.~(in preparation). Extracting
  relevant information.
    \item Brenner, N., Bialek, W., \& de Ruyter van Steveninck, R.\
  (2000). Adaptive rescaling optimizes information transmission, {\em
    Neuron} {\bf 26}, 695--702.
    \item Brunel, N., \& Nadal, P.~(1998). Mutual information, Fisher
  information, and population coding, {\em Neural Comp.} {\bf 10},
  1731-1757.
    \item Bruder, S.~D.~(1998).  Ph.D.~Dissertation, Princeton
  University.
    \item Capaldi, E.~A., ~Robinson, G.~E., \& Fahrbach S.~E.~(1999).
  Neuroethology of spatial learning: The birds and the bees, {\em
    Annu.~Rev.~Psychol.} {\bf 50}, 651--682.
    \item Cesa--Bianchi, N., \& Lugosi, G.~(2000). Worst-case bounds
  for the logarithmic loss of predictors, to appear in {\em Machine
    Learning}.
    \item Carpenter, R.~H.~S., \& Williams, M.~L.~L.~(1995). Neural
  computation of log likelihood in control of saccadic eye movements,
  {\em Nature} {\bf 377}, 59--62.
    \item Chaitin, G.~J.~(1975). A theory of program size formally
  identical to information theory, {\em J.~Assoc.~Comp.~Mach.}  {\bf
    22}, 329--340.
    \item Clarke, B.~S., \& Barron A.~R.~(1990).
  Information--theoretic asymptotics of Bayes methods, {\em IEEE
    Trans.~Inf.~Thy.} {\bf 36}, 453--471.
    \item Cover, T.~M., \& King, R.~C.~(1978).  A convergent gambling
  estimate of the entropy of English, {\em IEEE Trans.~Inf.~Thy.} {\bf
    24}, 413--421.
    \item Cover, T.~M., \& Thomas, J.~A.~(1991).  {\em Elements of
    Information Theory} (Wiley, New York).
    \item Crutchfield, J.~P., \& Feldman, D.~P.~(1997). Statistical
  complexity of simple 1--d spin systems, {\em Phys.~Rev.~E} {\bf 55},
  1239--1243R. See also cond-mat/9702191.
    \item Crutchfield, J.~P., Feldman, D.~P., \& Shalizi C.~R.~(1999).
  Comment on ``Simple measure for complexity,'' preprint.  Available
  at chao-dyn/9907001.
    \item Crutchfield, J.~P., \& Shalizi, C.~R.~(1998). Thermodynamic
  depth of causal states: objective complexity via minimal
  representation, {\em Phys.~Rev.~E} {\bf 59}, 275--283.  See also
  cond-mat/9808147.
    \item Crutchfield, J.~P., \& Young, K.~(1989). Inferring
  statistical complexity, {\em Phys.\ Rev.\ Lett.} {\bf 63}, 105--108.
    \item Eagleman, D.~M., \& Sejnowski, T.~J.\ (2000).  Motion
  integration and postdiction in visual awareness, {\em Science} {\bf
    287,} 2036--2038.
    \item Ebeling, W., \& P{\"o}schel, T.~(1994). Entropy and
  long-range correlations in literary English, {\em Europhys.~Lett.}
  {\bf 26}, 241--246.
    \item Feldman, D.~P., \& Crutchfield, J.~P.~(1998). Measures of
  statistical complexity: why? {\em Phys.~Lett.~A} {\bf 238},
  244--252.  See also cond-mat/9708186.
    \item Gell--Mann, M., \& Lloyd, S.~(1996). Information measures,
  effective complexity, and total information, {\it Complexity} {\bf
    2}, 44--52.
    \item de Gennes, P.--G.~(1979). {\em Scaling Concepts in Polymer
    Physics} (Cornell University Press, Ithaca and London).
    \item Grassberger, P.~(1986).  Toward a quantitative theory of
  self--generated complexity, {\it Int.~J.~Theor.~Phys.} {\bf 25},
  907--938.
    \item Grassberger, P.~(1991). Information and complexity measures
  in dynamical systems, in {\em Information Dynamics}, H.\ 
  Atmanspacher and H.\ Schreingraber, eds., pp.~15--33 (Plenum
  Press, New York).
    \item Hall, P., \& Hannan E.~(1988). On stochastic complexity and
  nonparametric density estimation, {\em Biometrica} {\bf 75},
  705--714.
    \item Haussler, D., Kearns, M., Seung, S., \& Tishby, N.~(1996).
  Rigorous learning curve bounds from statistical mechanics, {\em
    Machine Learning} {\bf 25}, 195--236.
    \item Haussler, D., Kearns, M., \& Schapire, R.~(1994). Bounds on
  the sample complexity of Bayesian learning using information theory
  and the VC dimension, {\it Machine Learning}, {\bf 14}, 83--114.
    \item Haussler, D., \& Opper, M.~(1995). General bounds on the
  mutual information between a parameter and {\em n} conditionally
  independent events. In {\em Proceedings of the Eighth Annual
    Conference on Computational Learning Theory}, 402--411 (ACM Press,
  New York).
    \item Haussler, D., \& Opper, M.~(1997). Mutual information,
  metric entropy and cumulative relative entropy risk, {\em Ann.\ 
    Statist.}, {\bf 25}, 2451--2492.
    \item Haussler, D., \& Opper, M.~(1998). Worst case prediction
  over sequences under log loss, in {\em The Mathematics of
    Information Coding, Extraction and Distribution}, G.\ Cybenko, D.\ 
  O'Leary and J.\ Rissanen, eds.
  (Springer--Verlag, New York). 
    \item Hawken, M.~J., \& Parker, A.~J.~(1991). Spatial receptive
  field organization in monkey V1 and its relationship to the cone
  mosaic, in {\it Computational models of visual processing}, M.~S.
  Landy and J.~A.~ Movshon, eds., pp.~83--93 (MIT Press, Cambridge)
    \item Herschkowitz, D., \& Nadal, J.--P.~(1999). Unsupervised and
  supervised learning: mutual information between parameters and
  observations, {\it Phys.~Rev.~E}, {\bf 59}, 3344--3360.
    \item Hilberg, W.~(1990). The well--known lower bound of
  information in written language: Is it a misinterpretation of
  Shannon experiments? (in German) {\it Frequenz} {\bf 44}, 243--248.
    \item Holy, T.~E.~(1997).  Analysis of data from continuous
  probability distributions, {\em Phys.~Rev.~Lett.}  {\bf 79},
  3545--3548.  See also physics/9706015.
    \item Ibragimov, I.\ \& Hasminskii, R.~(1972). On an information
  in a sample about a parameter, in {\it Second International
    Symposium on Information Theory}, 295-309 (IEEE Press, New York).
    \item Kang, K., \& Sompolinsky, H.~(2001). Mutual information of
  population codes and distance measures in probability space,
  preprint. Available at cond-mat/0101161.
    \item Kemeney, J.~G.~(1953).  The use of simplicity in induction,
  {\em Philos.~Rev.} {\bf 62}, 391--315.
    \item Kolmogoroff, A.~(1939).~ Sur l'interpolation et
  extrapolations des suites stationnaires, {\em C.~R.~Acad.~Sci.
    Paris} {\bf 208}, 2043--2045.
    \item Kolmogorov, A.~N.~(1941).  Interpolation and extrapolation
  of stationary random sequences (in Russian), in {\em Izv.\ Akad.\ 
    Nauk.\ SSSR Ser.\ Mat.} {\bf 5}, 3--14; translation in {\em
    Selected Works of A.~N.~Kolmogorov, vol.~II.}  A.~N.~Shirya\-gev,
  ed., pp.~272--280 (Kluwer Academic Publishers, Dordrecht, The
  Netherlands).
    \item Kolmogorov, A.~N.~(1965).  Three approaches to the
  quantitative definition of information, {\em Prob.~Inf.~Trans.}
  {\bf 1}, 4--7.
    \item Li, M., \& Vit{\'a}nyi, P.~(1993). {\em An Introduction to
    Kolmogorov Complexity and its Applications} (Springer--Verlag, New
  York).
    \item Lloyd, S., \& Pagels, H.~(1988). Complexity as thermodynamic
  depth, {\em Ann.\ Phys.} {\bf 188}, 186--213.
    \item Logothetis, N.~K., \& Sheinberg, D.~L.~(1996).  Visual
  object recognition, {\em Annu.\ Rev.\ Neurosci.} {\bf 19}, 577--621.
    \item Lopes, L.~L., \& Oden, G.~C.~(1987). Distinguishing between
  random and nonrandom events, {\em J.~Exp.~Psych.: Learning, Memory,
    and Cognition} {\bf 13}, 392--400.
    \item Lopez--Ruiz, R., Mancini, H.~L., \& Calbet, X.~(1995). A
  statistical measure of complexity, {\em Phys.~Lett.~A} {\bf 209},
  321--326.
    \item MacKay, D.~J.~C.~(1992).  Bayesian interpolation, {\em
    Neural Comp.} {\bf 4}, 415--447.
    \item Nemenman, I.\ (2000). Information theory and learning: a
  physical approach, Ph.~D.~Dissertation.  Princeton University. Sea
  also physics/0009032.
    \item Nemenman, I., \& Bialek, W.\ (2001).  Learning continuous
  distributions: Simulations with field theoretic priors.  To appear
  in {\em Advances in Neural Information Processing Systems} {\bf 13},
  T.\ K.\ Leen, T.\ G.\ Dietterich, and V.\ Tresp, eds. (MIT Press,
  Cambridge). See also cond-mat/0009165.
    \item Opper, M.~(1994). Learning and generalization in a
  two--layer neural network: the role of the Vapnik--Chervonenkis
  dimension, {\it Phys.~Rev.~Lett.} {\bf 72}, 2113--2116.
    \item Opper, M., \& Haussler, D.~(1995). Bounds for predictive
  errors in the statistical mechanics of supervised learning, {\it
    Phys.~Rev.~Lett.}  {\bf 75}, 3772--3775.
    \item Periwal, V.~(1997).  Reparametrization invariant statistical
  inference and gravity, {\em Phys.~Rev.~Lett.} {\bf 78}, 4671--4674.
  See also hep-th/9703135.
    \item Periwal, V.~(1998).  Geometrical statistical inference, {\em
    Nucl.~Phys.~B} {\bf 554[FS]}, 719--730.  See also
  adap-org/9801001.
    \item P\"oschel, T., Ebeling, W., \& Ros\'e, H.~(1995). Guessing
  probability distributions from small samples, {\it J.~Stat.~Phys.}
  {\bf 80}, 1443--1452.
    \item Reinagel, P., \& Reid, R.\ C.\ (2000).  Temporal coding of
  visual information in the thalamus, {\em J.\  Neurosci.} {\bf 20,}
  5392--5400.
    \item Renyi, A (1964) On the amount of information concerning an
  unknown parameter in a sequence of observations. {\it Publ.\ Math.\ 
    Inst.\ Hungar.\ Acad.\ Sci.}, {\bf 9}, 617--625.
    \item Rieke, F., Warland, D., de Ruyter van Steveninck, R., \&
  Bialek, W.~(1997).  {\em Spikes: Exploring the Neural Code} (MIT
  Press, Cambridge).
    \item Rissanen, J.~(1978).  Modeling by shortest data description,
  {\em Automatica}, {\bf 14}, 465--471.
    \item Rissanen, J.~(1984). Universal coding, information,
  prediction, and estimation, {\em IEEE Trans.~Inf.~Thy.}  {\bf 30},
  629--636.
    \item Rissanen, J.~(1986). Stochastic complexity and modeling,
  {\em Ann.~Statist.} {\bf 14}, 1080--1100.
    \item Rissanen, J.~(1987). Stochastic complexity, {\em J.~Roy.
    Stat.~Soc.~B}, {\bf 49}, 223--239, 253--265.
    \item Rissanen, J.~(1989). {\em Stochastic Complexity and
    Statistical Inquiry} (World Scientific, Singapore).
    \item Rissanen, J.~(1996).  Fisher information and stochastic
  complexity, {\em IEEE Trans.\ Inf.\ Thy.} {\bf 42}, 40--47.
    \item Rissanen, J., Speed, \& T., Yu, B.~(1992). Density
  estimation by stochastic complexity, {\em IEEE Trans.\ Inf.\ Thy.}
  {\bf 38}, 315--323.
    \item Saffran, J.~R., Aslin, R.~N., \& Newport, E.~L.~(1996).
  Statistical learning by 8--month--old infants, {\em Science} {\bf
    274}, 1926--1928.
    \item Saffran, J.~R., Johnson, E.~K., Aslin, R.~H., \& Newport, E.
  L.~(1999).  Statistical learning of tone sequences by human infants
  and adults, {\em Cognition} {\bf 70}, 27--52.
    \item Seung, H.~S., Sompolinsky, H., \& Tishby, N.~(1992).
  Statistical mechanics of learning from examples, {\em Phys.~Rev.~A}
  {\bf 45}, 6056--6091.
    \item Shalizi, C.~R., \& Crutchfield, J.~P.~(1999). Computational
  mechanics: pattern and prediction, structure and simplicity,
  preprint.  Available at \\cond-mat/9907176.
    \item Shalizi, C.~R., \& Crutchfield, J.~P.~(2000). Information
  bottleneck, causal states, and statistical relevance bases: how to
  to represent relevant information in memoryless transduction.
  Available at nlin.AO/0006025.
    \item Shiner, J., Davison, M., \& Landsberger, P.~(1999). Simple
  measure for complexity, {\em Phys.~Rev.~E} {\bf 59}, 1459--1464.
    \item Shannon, C.~E.~(1948).  A mathematical theory of
  communication, {\em Bell Sys.\ Tech.\ J.} {\bf 27}, 379--423,
  623--656.  Reprinted in C.~E.~Shannon and W.~Wea\-ver, {\em The
    Mathematical Theory of Communication} (University of Illinois
  Press, Urbana, 1949).
    \item Shannon, C.~E.~(1951).  Prediction and entropy of printed
  English, {\em Bell Sys.\ Tech.\ J.} {\bf 30}, 50--64. Reprinted in
  N.~J.~A.~Sloane and A.~D.~Wyner, eds., {\em Claude Elwood Shannon:
    Collected papers} (IEEE Press, New York, 1993).
    \item Smirnakis, S., Berry III, M.~J., Warland, D.~K., Bialek, W.,
  \& Meister, M.\ (1997).  Adaptation of retinal processing to image
  contrast and spatial scale, {\em Nature} {\bf 386}, 69--73.
    \item Sole, R.~V., \& Luque, B.~(1999).  Statistical measures of
  complexity for strongly interacting systems, preprint.  Available at
  adap-org/9909002.
    \item Solomonoff, R.~J.~(1964).  A formal theory of inductive
  inference, {\em Inform.~and Control} {\bf 7}, 1--22, 224--254.
    \item Strong, S.\ P., Koberle, R., de Ruyter van Steveninck, R.,
  \& Bialek, W.\ (1998). Entropy and information in neural spike
  trains, {\it Phys.\ Rev.\ Lett.} {\bf 80,} 197--200.
    \item Tishby, N., Pereira, F., \& Bialek, W.~(1999).  The
  information bottleneck method, in {\em Proceedings of the 37th
    Annual Allerton Conference on Communication, Control and
    Computing}, B.~Hajek and R.~S.~Sreenivas, eds., pp.~368--377
  (University of Illinois). See also physics/0004057
    \item Vapnik, V.~(1998). \emph{Statistical Learning Theory} (John
  Wiley \& Sons, New York).
    \item Vit{\'a}nyi, P., \& Li, M.~(2000). Minimum description
  length induction, Baye\-sianism, and Kolmogorov Complexity, {\em IEEE
    Trans.~Inf.~Thy.}  {\bf 46}, 446--464. See also cs.LG/9901014.
    \item Weigend, A.~S., \& Gershenfeld, N.~A., eds.\ (1994). {\em
    Time series prediction: Forecasting the future and understanding
    the past} (Addison--Wesley, Reading MA).
    \item Wiener, N.~(1949).  {\em Extrapolation, Interpolation and
    Smoothing of Time Series} (Wiley, New York).
    \item Wolfram, S.~(1984). Computation theory of cellular automata,
  {\em Commun.\ Math.\ Physics} {\bf 96}, 15--57.
    \item Wong, W., \& Shen, X.~(1995) Probability inequalities for
  likelihood ratios and convergence rates of sieve MLE's, {\em Ann.\ 
    Statist.} {\bf 23}, 339--362. 
\end{description}
\end{document}